\documentclass[a4paper,11pt]{article}
\pdfoutput=1 

\usepackage{jcappub} 

\usepackage[T1]{fontenc} 
\usepackage[dvipsnames]{xcolor}
\usepackage{tikz}

\usepackage{amssymb, amsmath, epsfig, natbib}

\renewcommand{\vec}[1]{\mathbf{#1}}

\newcommand{\vq}{\vec{q}}
\newcommand{\vx}{\vec{x}}
\newcommand{\vk}{\vec{k}}
\newcommand{\vPsi}{\vec{\Psi}}

\newcommand{\df}{\delta}
\newcommand*{\non}  {\nonumber}
\newcommand*{\lb}  {\left(}
\newcommand*{\rb}  {\right)}

\newcommand*{\la}  {\left\langle}
\newcommand*{\ra}  {\right\rangle}

\author[a,b]{Zvonimir Vlah}
\author[c]{Emanuele Castorina}
\author[c,d]{Martin White}

\affiliation[a]{Stanford Institute for Theoretical Physics and Department of Physics, Stanford University, Stanford, CA 94306, USA} 
\affiliation[b]{Kavli Institute for Particle Astrophysics and Cosmology, SLAC and Stanford University, Menlo Park, CA 94025, USA}
\affiliation[c]{Department of Physics, University of California,
Berkeley, CA 94720}
\affiliation[d]{Department of Astronomy, University of California,
Berkeley, CA 94720}

\emailAdd{zvlah@stanford.edu}
\emailAdd{ecastorina@berkeley.edu}
\emailAdd{mwhite@berkeley.edu}

\title{The Gaussian streaming model and Convolution Lagrangian
effective field theory}


\abstract{We update the ingredients of the Gaussian streaming model (GSM) for
the redshift-space clustering of biased tracers using the techniques of
Lagrangian perturbation theory, effective field theory (EFT) and a generalized
Lagrangian bias expansion.
After relating the GSM to the cumulant expansion, we present new results for
the real-space correlation function, mean pairwise velocity and pairwise
velocity dispersion including counter terms from EFT and bias terms through
third order in the linear density, its leading derivatives and its shear up
to second order.
We discuss the connection to the Gaussian peaks formalism.
We compare the ingredients of the GSM to a suite of large N-body simulations,
and show the performance of the theory on the low order multipoles of the
redshift-space correlation function and power spectrum.
We highlight the importance of a general biasing scheme, which we find to be
as important as higher-order corrections due to non-linear evolution for
the halos we consider on the scales of interest to us.
}
\arxivnumber{1609.02908}

\begin{document}
\maketitle
\flushbottom

\section{Introduction}

The growth of the large-scale structure in the observed Universe
arises due to gravitational collapse into dark-matter dominated
potentials, tempered by the expansion of the Universe.
A wealth of information can be encoded in the growth rate,
including constraints on the expansion history, the nature of dark
energy and modified gravity \cite{Wei13,PDG14}.
There are several ways of studying the growth of structure, but perhaps
the oldest and highest signal-to-noise measurement comes from observations
of the anisotropy in the clustering of objects in redshift surveys.
Since the redshift, from which one infers distance, contains a contribution
from the line-of-sight velocity the clustering of objects in redshift surveys
exhibits anisotropy \cite{Kai87,Ham98,Wei13}, known as redshift-space
distortions (RSD).

Over the last several decades measurements of the redshift-space clustering
of galaxies have become increasingly precise
(for example the measurements from the recently completed BOSS survey
\cite{Daw13} have percent-level uncertainties on quasi-linear scales)
allowing highly precise tests of the current paradigm but also posing a
significant challenge for theorists seeking to model the data.
As we prepare for the next generation of surveys we need models which are
able to model the redshift-space clustering of biased tracers at the percent
level over a wide range of scales (to break degeneracies between parameters).

In this paper we shall attempt to model the low-order moments of the
redshift-space clustering signal in configuration space, using models
based upon Lagrangian perturbation theory (LPT; see below) and the
effective field theory of large-scale structure (EFT
\cite{BNSZ12,CHS12,PajZal13,Man14,MerPaj14,CLP14,PorSenZal14,VlaWhiAvi15}).
LPT is one of the oldest and most successful analytic model for studying
large-scale structure, and provides a simple connection to N-body simulations
and peaks theory \cite{BBKS}.
EFT is a consistent method for incorporating the effects of non-perturbative
physics into perturbation theory by including a number of additional terms,
with free parameters, whose structure is determined by the symmetries of the
theory.
Our focus here will be on increasing the precision with which we can predict
the clustering moments on intermediate scales ($>25\,h^{-1}$Mpc), rather
than on increasing the range of scales we predict.  We believe this is a
more appropriate use of techniques built upon perturbation theory.
For an alternative route, see Ref.~\cite{Rei14}.

The outline of this paper is as follows.
In the next section we review the context within which we will do our
calculations: (Lagrangian) perturbation theory and the (Gaussian) streaming
model.  The next section describes the computation of each piece of the
streaming model in terms of perturbation theory, including our bias model and
the impact of small-scale physics which is not explicitly modeled.
In section \ref{sec:evolution} we discuss the expected evolution of the
correlation function, and the extent to which measurements in finite redshift
slice can be interpreted as being at an ``effective'' redshift.
We present some preliminary comparison of our results with N-body simulations
in section \ref{sec:nbody}, finding that we are limited by the accuracy of the
simulations.
Finally we conclude in section \ref{sec:conclusions}.
A number of technical steps are relegated to a series of appendices.
Appendix \ref{app:GSM} reviews the derivation of the streaming model.
Appendix \ref{app:fourier} reviews the Fourier-space statistics.
Appendix \ref{app:derivs} details the computation of the time-derivative
terms which enter in the velocity statistics.
Appendix \ref{app:bias} presents a more general Lagrangian bias model which
includes derivatives of the initial density field and shear terms.
Appendix \ref{app:DF} compares our formalism to the distribution function
formalism of Ref.~\cite{SelMcD11}.

\section{Background}
\label{sec:background}

The next few subsections present some background to set the stage and our
notation.  First we review LPT and then the ``streaming model''.  Some of
the derivations or more technical details are relegated to appendices.

Our focus will be the prediction of the two-point function in redshift space,
i.e.~the correlation function:
\begin{equation}
  \xi(\vec{s}) = \langle \delta(\vec{x}) \delta(\vec{x}+\vec{s}) \rangle.
\end{equation}
which depends on the separation, $s=|\mathbf{s}|$, and the cosine of the
angle between the separation and the line-of-sight\footnote{We adopt the
``plane-parallel'' approximation throughout, so that the line-of-sight (LOS)
is chosen along a single Cartesian axis: $\hat{\vec{z}}$.}
$\mu\equiv\hat{\vec{s}}\cdot\hat{\vec{z}}$.
Following common practice we project the $\mu$ dependence onto Legendre
polynomials, $\mathcal{P}_\ell$,
\begin{equation}
  \xi(s,\mu) = \sum_{\ell} \xi_{\ell}(s) \mathcal{P}_{\ell}(\mu)\ .
\label{eqn:legendre-expansion}
\end{equation}
By symmetry, odd $\ell$ moments vanish.
In linear theory, only $\ell= 0,2,4$ contribute; we will focus our
model predictions on those moments.  

\subsection{Lagrangian perturbation theory}

The Lagrangian approach to cosmological structure formation was developed
in \cite{Zel70,Buc89,Mou91,Hiv95,TayHam96,Mat08a,Mat08b,CLPT,Mat15} and
traces the trajectory of an individual fluid element through space and time.
For a fluid element located at position $\vq$ at some initial time $t_0$,
its position at subsequent times can be written in terms of the Lagrangian
displacement field $\vPsi$,
\begin{equation}
    \vx(\vq,t) = \vq + \vPsi(\vq,t) ,
\label{eqn:LtoE}
\end{equation}
where $\vPsi(\vq,t_0) = 0$.  Every element of the fluid is uniquely labeled by
$\vq$ and $\vPsi(\vq,t)$ fully specifies the evolution.  In what follows we
shall suppress the time dependence of $\vPsi$ for notational convenience.
Once $\vPsi(\vq)$ is known, the density field at any time is simply
\begin{equation}
  1+\delta(\vx)=\int d^3q\ \delta_D\big[\vx-\vq-\vPsi(\vq)\big]
  \quad\Rightarrow\quad
  \delta(\vk)=\int d^3q\ e^{i\vk\cdot\vq}\big(e^{i\vk\cdot\vPsi(\vq)}-1\big)
\label{eqn:deltadefn}
\end{equation}
For tracers which are biased, the density is modulated by a function
of the linear density field, the Laplacian of the linear density field
and shear field at $\mathbf{q}$ (see \S\ref{sec:ingredients}).
The evolution of $\vPsi$ is governed by
$\partial_t^2\vPsi + 2H\partial_t\vPsi = -\nabla\Phi(\vq+\vPsi)$.
We shall work throughout in terms of conformal time, $d\eta=dt/a$, and
write $\mathcal{H}=aH$ for the conformal Hubble parameter.  The equation of
motion is thus
$\ddot{\vPsi} + \mathcal{H}\dot{\vPsi} = -\nabla\Phi(\vq+\vPsi)$,
where overdots indicate derivatives w.r.t.~conformal time.
In LPT one finds a perturbative solution for $\vPsi$,
$\vPsi = \vPsi^{(1)} + \vPsi^{(2)} + \vPsi^{(3)} + \cdots$,
with the first order solution, linear in the density field, being the
Zeldovich approximation \cite{Zel70}.
Higher order solutions are specified in terms of integrals of higher powers
of the linear density field \cite{ZheFri14,Mat15}.
To these perturbative terms are then added a series of `extra' terms, which
encapsulate the effect of the small-scale physics which is missing in the
perturbative treatment \cite{PorSenZal14,VlaWhiAvi15}.

\subsection{The streaming model}

The effects of super-cluster infall \cite{Kai87} on large scales, and
virial motions within clusters on small scales, act to modify the clustering
pattern observed in redshift space.  Early models of this effect
\cite{Pea92,Par94,PD94,BPH96,Ham98,HatCol99},
nowadays termed ``dispersion models'', were primarily phenomenological in
nature and treated the two regimes independently.  While successful at
describing early survey data they do not describe the effects at the level
of detail necessary for current and future surveys.

A closely related class of models, inspired by \cite{Pee80,Fis95},
are streaming models which aim to jointly model the density and velocity field.
The number of pairs of objects in real space is related to $1+\xi$, so if one
has a model for the probability, $\mathcal{P}$, that a pair with real-space,
line-of-sight separation $r_{\parallel}$ will be observed with redshift-space,
line-of-sight separation $s_{\parallel}$ then pair conservation implies
\begin{equation}
  1+\xi^s(s_\perp,s_\parallel) = \int dy\ \left[1+\xi(\mathbf{r})\right]
  \mathcal{P}\left(y=s_\parallel-r_\parallel|\mathbf{r}\right)
\end{equation}
Our focus will be on the Gaussian streaming model (GSM), as originally
developed in \cite{ReiWhi11,Rei12,WanReiWhi14} and discussed in
\cite{BiaChiGuz15,Bia16,Uhl15,Kop16}.
In the GSM we assume that -- for massive enough halos at sufficiently high
redshifts -- $\mathcal{P}$ can be well approximated by a Gaussian, so that
the redshift-space correlation function can be written
(see Appendix \ref{app:GSM})
\begin{equation}
  1+\xi^s(s_\perp,s_\parallel) = \int\frac{dy}{\sqrt{2\pi}\ \sigma_{12}}
  \left[1+\xi\right]
  \exp\left\{-\frac{[s_\parallel-y-\mu v_{12}]^2}{2\sigma_{12}^2}\right\}
\label{eqn:gsm}
\end{equation}
with $r=\sqrt{y^2+s_\perp^2}$,
$\mu=s_\parallel/\sqrt{s_\perp^2+s_\parallel^2}$,
$\xi$ the real-space correlation function (of the biased tracer),
$v_{12}$ the mean, pairwise infall velocity and
$\sigma_{12}$ the pairwise dispersion.
The correlation function, $\xi$, and the velocity moments
($v_{12}$ and $\sigma_{12}$) are functions of $r$, though that dependence
has been suppressed in Eq.~(\ref{eqn:gsm}).

An alternative asymptotic expansion to the GSM is the ``Edgeworth streaming
model'' of Ref.~\cite{Uhl15}.  This keeps higher orders in the cumulant
expansion (e.g.~Appendix \ref{app:GSM}) which improves the performance of
the model at smaller scales and for higher multipoles.
Ref.~\cite{Uhl15} show that for the halo masses they probed innaccuracies
in the perturbative expansion are more significant than the neglect of the
higher-order cumulants in the GSM, suggesting that improving the ingredients
in the GSM will yield the most benefit.  This mirrors the conclusions of
Ref.~\cite{ReiWhi11}, and forms part of the motivation for this work.

Ref.~\cite{Uhl15} also explore a number of coarse-graining approaches for
perturbative evaluation of the pieces of the GSM, an approach which has been
furthered by Ref.~\cite{Kop16}.
As we will discuss later, smoothing the initial field can mimic some of the
effects of EFT operators (e.g.~it can reduce the rms displacement, which tends
to be over-predicted by LPT) and hence improve the numerical agreement between
perturbation theory and simulations.  However there are couplings between the
long- and short-scale modes which need to be ``integrated out'' which are not
taken into account in the smoothing approach.  The apparent need for
different smoothing scales in different cumulants can be naturally explained
by the structure of the EFT counter terms entering the calculation.

\section{The model ingredients} 
\label{sec:ingredients}


In this section we review the predictions of ``Lagrangian effective field
theory'' for the ingredients going into the streaming model described above.
Much of this material has been well developed in the literature, so we
shall present here only the main results.
The reader is referred to
Refs.~\cite{Mat08a,Mat08b,CLPT,WanReiWhi14,VlaSelBal15} for the development
of `standard' LPT and Refs.~\cite{PorSenZal14,VlaWhiAvi15} for the effective
field theory corrections.
We shall closely follow the notation in Refs.~\cite{Mat08a,WanReiWhi14}, in
particular we define
$\langle\delta(\mathbf{q}_1)\delta(\mathbf{q}_2)\rangle=
 \langle\delta_1\delta_2\rangle_c = \xi(\mathbf{q}=\mathbf{q}_1-\mathbf{q}_2)$,
$\Delta_i=\Psi_i(\mathbf{q}_1)-\Psi_i(\mathbf{q}_2)$ and
\begin{equation}
  U_i^{mn} = \langle\delta_1^m\delta_2^n\Delta_i\rangle_c
  \quad , \quad
  A_{ij}^{mn} = \langle\delta_1^m\delta_2^n\Delta_i\Delta_j\rangle_c
  \quad , \quad
  W_{ijk} = \langle\Delta_i\Delta_j\Delta_k\rangle_c
\end{equation}
with the shorthand notation $A_{ij}=A_{ij}^{00}$ and $U_i = U_i^{10}$.

All of the LPT results presented in the paper rely on an approximation
where these kernels are time independent
(as in an Einstein-de Sitter cosmology).
This is a potential source of systematic error since in other cosmologies,
including $\Lambda$CDM, the LPT kernels can be time dependent.
The effect of this on the one-loop matter power spectrum is typically at
a sub-percent level \cite{Ber02,Takahashi0806,Fas2016}, but can reach 1\%
or more when also considering momentum statistics \cite{Fas2016} 
i.e.~pairwise velocities and velocity dispersions.
These effects are thus of particular interest if sub-percent predictions
of the statistics in redshift space are required, but will not dominate
our error budget.

We shall present our description of biased tracers in Lagrangian space.
We use a generalised approach where, in addition to the expansion in powers
of the linear density field $\df_L$ (Refs.~\cite{Mat08a,Mat08b}), we include
the shear of this field $s_{ij}$ up to second order
(see e.g. \cite{Cat98,Cat00,McDRoy09,Des11,Mus12,Cha12,
Bal12,SchJeoDes13,Par13b,Whi14,MatDes16})
and an explicit derivative expansion starting with
$(\nabla^2/\Lambda_L^2)\df_L$ \cite{McDRoy09,Ass14,Sen14,Mirb14,Ang15,Fujita2016}.
The associated scale, $1/\Lambda_L$, is related to the typical size of
the object we are describing (also called also proto-halos).  We shall refer
to it as the Lagrangian radius of the halo.
Thus we can write
\begin{equation}
  1+\delta_X(\vx)=\int d^3q
  \ F\big[\delta(\vq),\nabla^2\delta(\vq),s^2(\vq)\big]
  \delta_D\big[\vx-\vq-\vPsi(\vq)\big]
\label{eqn:deltaX}
\end{equation}
with $s^2=s_{ij}s_{ij}$.
For more detailed description of the biasing model we refer the reader
to Appendix \ref{app:bias}.
After auto-correlating the field we keep the biasing terms up to quadratic
order (in the bias expansion) and one loop in the perturbative description
of the nonlinear dynamics.
We denote the first two functional derivative of $F$ w.r.t~$\delta$ as
$b_1$ and $b_2$, the functional derivative w.r.t.~$\nabla^2\delta$ as
$b_{\nabla^2}$ and the functional derivative w.r.t.~$s^2$ as $b_{s^2}$. 

The corresponding terms of the general expansion in the Eulerian framework
have been discussed in \cite{McDRoy09,Ass14,Sen14,Mirb14,Ang15} who also
include third order terms.
The precise relation of these Eulerian and Lagrangian biasing terms is
not simple and parts of the third order Eulerian terms are dynamically
produced from evolving the Lagrangian bias framework
\cite{Cat98,Cat00,Cha12,Bal12,Saito14,Biag2014}.
We intend to return to this question in a future publication.

Alternatively, biasing of dark matter halos can be studied in a
phenomenological framework like the excursion set model \cite{Mus12} 
or the theory of Gaussian peaks \cite{BBKS,Des08,Par13a}.
The derivative (or in general `scale dependent bias') and tidal terms
can be analytically derived within these frameworks leading to predictions
for the corresponding bias coefficients.  For example, the derivative terms
arise in the peaks model as a consequence of the peak constraint imposed on
the initial field when smoothed on the halo scale \cite{Par13b,Sheth13,Bal15a}.
These predictions can be very useful in describing the time evolution and
approximate values of these bias parameters, which can be used
e.g.~as priors in the EFT framework (see \S\ref{sec:bias_philosophy}).

Finally, we note that the one-loop integrals that appear in the calculation
of both the LPT and bias terms typically rely on the computation of 2D numerical
integrals (with higher loops involving higher dimensional integration). 
It has been shown recently that all of these integrals can be recast in
a convenient form where Hankel transforms can be used to speed up the
computation (see e.g.~Appendix B in Ref.~\cite{Schmittfull2016} for the
one-loop LPT expressions and Ref.~\cite{Schmittfull2016v2} for discussion of
bias and RSD terms).  Some of our biasing terms are already expressed in
this form (see Appendix \ref{app:bias}).

\subsection{The real-space correlation function}
\label{sec:xir}

The real-space correlation function for tracers which are locally biased
in Lagrangian space is given as the integral
\begin{equation}
  1 + \xi(r) = \int d^3q\ M_0(\vec{r},\vec{q})\ .
\label{eqn:xi-exp}
\end{equation}
where \cite{CLPT,WanReiWhi14,VlaSelBal15}
\begin{eqnarray}
    M_0 &=&
    \dfrac{1}{(2\pi)^{3/2}|A_{\rm lin}|^{1/2}}
    e^{-(1/2)(q_i-r_i)(A_{\rm lin}^{-1})_{ij}(q_j-r_j)}
    \nonumber \\
    &\times& \bigg\{ 1 - \dfrac{1}{2}G_{ij}A_{ij}^{1-{\rm loop}}
    + b_1^2 \xi_L + \dfrac{1}{2}b_2^2 \xi_L^2 - 2
    b_1 U_ig_i + \dfrac{1}{6} W_{ijk}\Gamma_{ijk}
    - [b_2 + b_1^2] U_i^{(1)}U_j^{(1)}G_{ij}
    \nonumber \\
    && - b_1^2U^{11}_ig_i - b_2 U^{20}_ig_i
    - 2 b_1 b_2 \xi_L U_i^{(1)}g_i - b_1 A^{10}_{ij}G_{ij}
    \nonumber \\
    &&- b_{s^2}\left(G_{ij}\Upsilon_{ij}+2g_iV_i^{10}\right) +
    b_{s^2}^2\zeta-2b_1b_{s^2}g_iV_i^{12} +
    b_2b_{s^2}\chi^{12}
    \nonumber \\
    && -\dfrac{1}{2} \alpha_\xi {\rm tr}G
    + 2 b_{\nabla^2} \mathcal{B}+ 2 b_{\nabla^2} b_1 \mathcal{B}_{2,i} g_i
    + \cdots \bigg\}\ .
\label{eqn:xi-kernel-m}
\end{eqnarray}
and we have defined
\begin{equation}
  \begin{split}
    & g_i = (A_{\rm lin}^{-1})_{ij}(q_j-r_j)
    \ ,\quad G_{ij} = (A_{\rm lin}^{-1})_{ij} - g_ig_j\ ,
    \\
    & \Gamma_{ijk} = (A_{\rm lin}^{-1})_{ij} g_k
    + (A_{\rm lin}^{-1})_{ki} g_j + (A_{\rm lin}^{-1})_{jk} g_i
    - g_ig_jg_k\ .
  \end{split}
\label{eqn:gaussian-intg-aux}
\end{equation}
Note our sign convention for $g_i$ follows that of Ref.~\cite{WanReiWhi14}
and differs from that of Ref.~\cite{VlaWhiAvi15} by a minus sign.  This
accounts for the difference in the sign of the $\Gamma_{ijk}$ terms
when comparing those two papers.
It can be useful to have Fourier-space expressions as well, we relegate
these to Appendix \ref{app:fourier}.

The first $2$ lines within the Eq.~\eqref{eqn:xi-kernel-m} are the predictions of standard
Lagrangian perturbation theory.
In the final line, the term depending on $\alpha$ is the ``EFT term'' which
encapsulates the effects of unmodeled, small-scale physics.
The numerical prefactor is conventional, and serves to standarzie the $k$-space
expression.
The $b_{\nabla^2}$ terms on the last line come from the $\nabla^2\delta$ bias,
while the $b_{s^2}$ terms on the third line are from the shear-dependence of
the bias.
Further details of the bias model, and expressions for $\mathcal{B}$, $V$,
$\chi$, $\zeta$ and $\Upsilon$ terms, are given in Appendix \ref{app:bias}.

\begin{figure}
\begin{center}
\resizebox{\textwidth}{!}{\includegraphics{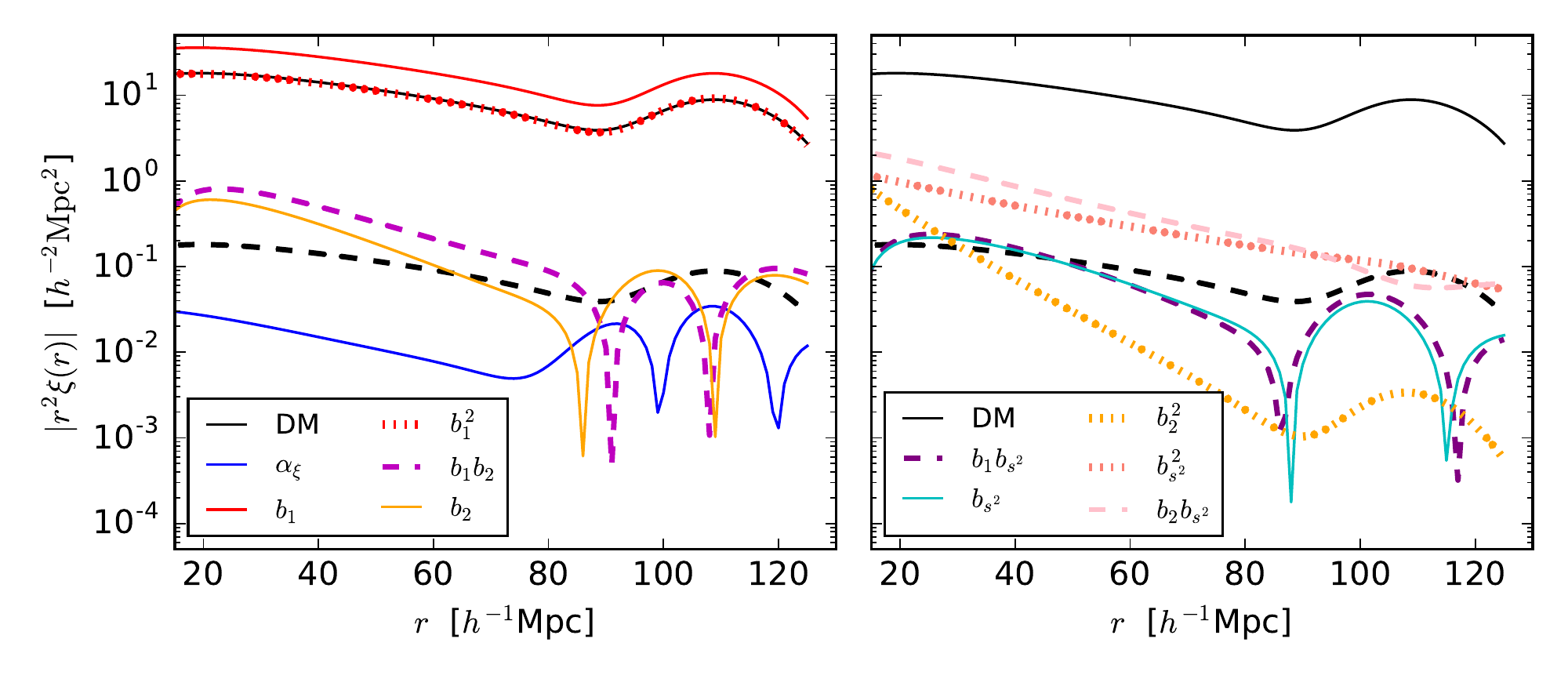}}
\end{center}
\caption{The terms in Eq.~(\ref{eqn:xi-kernel-m}) evaluated at $z=0.8$ for
illustration.  The left panel shows the terms independent of $b_{s^2}$ while
the right panel shows the $b_{s^2}$-dependent terms.  In each case the solid
black line shows the DM-only correlation function (times $r^2$) corresponding
to the ``1'' term in Eq.~(\ref{eqn:xi-kernel-m}).  To guide the eye on the
relative size of the other terms, 1 per cent of this is shown as the dashed
black line.  It is clear that to first approximation
$\xi(r)\approx(1+b_1)^2\xi_{\rm m}(r)$ with corrections coming in at the few
per cent level.  It is also clear that the shear terms are at least as
important as the $b_2$ terms.}
\label{fig:xir_pieces}
\end{figure}

The last three terms in Eq.~(\ref{eqn:xi-kernel-m}) all have leading order
contributions of the form $k^2\,P_L(k)$.
For this reason they are largely degenerate at the scales of interest and
we can fit for just one (combined) free parameter, which we take to be the
coefficient $\alpha_\xi$.
We emphasize that the degree of degeneracy is to some extent a numerical
coincidence, since PT is an expansion in $k/k_{NL}$ and bias is an expansion
in $k/\Lambda_L$.
The two scales are in principle separate, with $k_{NL}$ typically larger than
$\Lambda_L$.  For this reason higher order terms in $k/\Lambda_L$ could become
important at $k$'s for which the lowest order terms in $k/k_{NL}$ are
sufficient.  The nature and degree of degeneracies at higher order remains
an open question.  Also note that our resummation scheme (keeping the
$A_{\rm lin}$ in the exponent) breaks the degeneracies, but only by terms
which are higher order (see also the discussion in Ref.~\cite{VlaWhiAvi15};
where a similar argument is made to drop some of the other EFT counter terms
which would be degenerate with $k^2\,P_L(k)$ to lowest order --- similar
approximations, neglecting higher angular dependencies, are made in
\cite{SenZal15}).
We have verified that these differences are numerically small so henceforth
we shall neglect the $\mathcal{B}$ terms.

Ref.~\cite{WanReiWhi14} kept all of the 1-loop terms contributing to
$A_{ij}$ in the exponential, but here we have followed Ref.~\cite{VlaWhiAvi15}
and kept only the linear (Zeldovich) terms exponentiated while expanding
the 1-loop terms into the $\{\cdots\}$.
This is more consistent with our retention of only the lowest order EFT
terms (while keeping the IR modes resummed).
Note that our formalism allows for an alternative IR resummation procedure
that was described in Appendix B of Ref.~\cite{VlaWhiAvi15}, which is similar
to, albeit more general than, the procedure suggested in Ref.~\cite{SenZal15}.
Similarly we have kept only terms up to second order in the linear theory
power spectrum, $\mathcal{O}(P_L^2)$, in the $\{\cdots\}$ of
Eq.~\eqref{eqn:xi-kernel-m}.  
This implies that only the lowest order contributions to $U_i$ are used in
the terms indicated.  We have found that this makes little difference
numerically, but again is more self-consistent if we view the terms in
Eq.~\eqref{eqn:xi-kernel-m} as an expansion in powers of $P_L$.

Fig.~\ref{fig:xir_pieces} shows the contributions to $r^2\,\xi(r)$ evaluated
at $z=0.8$.  The left panel shows the terms independent of $b_{s^2}$ while
the right panel shows the $b_{s^2}$-dependent terms.  The dominant terms are
the ``1'', $b_1$ and $b_1^2$ terms whose sum closely approximates $(1+b_1)^2$
times the matter result.  The dashed black line in each panel shows 1\% of the
matter result as a guide to the level of correction.  We see that all of the
terms (with the possible exception of $b_2^2$) are important at the 1\% level.
Naively the $\alpha_\xi$ term is smaller than our 1\% mark, however we expect
$\alpha_\xi\sim\mathcal{O}(10)$ given our normalization convention so that
this term does become significant and can alter the width of the BAO peak (as
well as modifying the small-scale correlation function).

\subsection{The (mean) pairwise velocity}

The mean, pairwise velocity is $v_{12} = v_{12,n}\hat{r}_n$ with
\begin{equation}
  v_{12,n}(\vec{r}) = [1+\xi(r)]^{-1} \int d^3q\  M_{1,n}(\vec{r}, \vec{q})\ .
\label{eqn:v-exp}
\end{equation}
and \cite{WanReiWhi14}
\begin{align}
    M_{1,n} =& \dfrac{f}{(2\pi)^{3/2}|A_{\rm lin}|^{1/2}}
    e^{-(1/2)(q_i-r_i)(A_{\rm lin}^{-1})_{ij}(q_j-r_j)} \nonumber\\
    & \times \bigg\{ 2 b_1\dot{U}_n - g_i\dot{A}_{in} + b_2
    \dot{U}^{20}_n + b_1^2 \dot{U}^{11}_n
    -\dfrac{1}{2} G_{ij}\dot{W}_{ijn} - 2 b_1 g_i \dot{A}^{10}_{in}
     \nonumber \\
    & + 2 b_1b_2 \xi_L \dot{U}_n^{(1)}
      - [b_2+b_1^2]\left( g_i U_i^{(1)} \dot{U}_n^{(1)}
      +g_iU_i^{(1)}\dot{U}_n^{(1)}\right)
      - b_1^2 \xi_L g_i\dot{A}_{in}^{(1)}
      - 2b_1 G_{ij}U_i^{(1)}\dot{A}_{in}^{(1)}
    \nonumber \\
    & + \alpha_v \nabla_n\xi_L - \alpha_v' g_n 
       - 2 b_{\nabla^2} \mathcal{B}_{2,n} \nonumber \\
    & + b_{s^2}\left(2 \dot{V}^{10}_n-g_i \dot{\Upsilon}_{in}\right)
      +  b_1 b_{s^2} \dot{V}^{12}_n
      + \cdots \bigg\}\ ,
\label{eqn:v-kernel-m-2nd}
\end{align}
The dot notation and the relationship between the derivatives and base
quantities is elucidated in Appendix \ref{app:derivs}.

As above, the first $2$ lines within the Eq.~\eqref{eqn:v-kernel-m-2nd}
are the predictions of Lagrangian perturbation theory, while the terms
depending on $\alpha$ are the EFT terms which encapsulate the small-scale
physics and the extra terms are due to the scale-dependent and
shear-dependent bias.
As for $\xi$, we have truncated the expansion in Eq.~\eqref{eqn:v-kernel-m-2nd}
at second order in the linear theory power spectrum, $P_L$, which explains the
use of $U_i^{(1)}$ and $A_{ij}^{(1)}$ in some terms.

\subsection{The (pairwise) velocity dispersion }

We decompose the velocity dispersion tensor into components parallel to
and perpendicular to the separation.  The cumulant\footnote{The cumulant,
$\sigma^2$, was used in the original GSM of Ref.~\cite{ReiWhi11} and in
Ref.~\cite{Uhl15}.  The non-cumulant version, $\hat{\sigma}^2$, was used
in Ref.~\cite{WanReiWhi14}.
The differences manifest themselves primarily on small scales.
We shall use the cumulant throughout (see Appendix \ref{app:GSM}).}
which enters Eq.~(\ref{eqn:gsm}) can be written as
$\sigma_\parallel^2=\hat{\sigma}_\parallel^2-v_\parallel^2$ and
$\sigma_\perp^2=\hat{\sigma}_\perp^2$ where
\begin{equation}
  \hat{\sigma}^2_{\parallel} = \hat{\sigma}^2_{12,nm} \hat{r}_n \hat{r}_m
  \ ,\quad
  \hat{\sigma}^2_{\perp} = \frac{1}{2}\hat{\sigma}^2_{12,nm}
                           \left( \delta^K_{nm}-\hat{r}_n\hat{r}_m \right) \ .
\label{eqn:sig-components}
\end{equation}
with
\begin{equation}
  \hat{\sigma}^2_{12,nm}(\vec{r}) = [1+\xi(r)]^{-1}
           \int d^3q\  M_{2,nm}(\vec{r}, \vec{q})\ .
\label{eqn:sig-exp}
\end{equation}
and we have (see Ref.~\cite{WanReiWhi14} )
\begin{align}
    M_{2,nm} =& \dfrac{f^2}{(2\pi)^{3/2}|A_{\rm lin}|^{1/2}}
    e^{-(1/2)(q_i-r_i)(A_{\rm lin}^{-1})_{ij}(q_j-r_j)}
    \nonumber\\
    & \times \big\{ [b_1^2 + b_2 ]
    \left(\dot{U}_n^{(1)} \dot{U}_m^{(1)}
        + \dot{U}_n^{(1)}\dot{U}_m^{(1)}\right) - 2 b_1 \left(
    \dot{A}_{in}^{(1)}g_i\dot{U}_m^{(1)} +
    \dot{A}_{im}^{(1)}g_i\dot{U}_n^{(1)}\right)
    - \dot{A}_{im}^{(1)}\dot{A}_{jn}^{(1)}G_{ij}
    \nonumber\\
    & + \ddot{A}_{nm} + b_1^2 \xi_L\ddot{A}_{nm}^{(1)}
      - 2b_1 U_i^{(1)} g_i \ddot{A}_{nm}^{(1)} + 2b_1 \ddot{A}^{10}_{nm}
      + 2 b_{s^2} \ddot{\Upsilon}_{mn}-\ddot{W}_{inm}g_i
    \nonumber\\
    & + \alpha_\sigma \delta_{nm} + \beta_\sigma \delta_{nm} \xi_L
    + \ldots \big\}\ .
\label{eqn:sig-kernel-m-2nd}
\end{align}
where above the first two lines in brackets $\{\cdots\}$ are the
results from the usual Lagrangian perturbation theory, truncated to second
order in $P_L$, and the terms in the last line are the EFT caunterterms. 
The first, constant, counterterm is a ``contact'' or ``zero lag'' term
coming from all of the terms with a non-trivial $q\to\infty$ limit.
In Fourier space these terms require an integration over all modes, including
very high $k$ modes.  The second term, $\delta_{nm}\xi_L$, stands in for a
number of terms which cancel the high-$k$ sensitivity of the $\ddot{A}_{nm}$,
$\ddot{A}_{nm}^{10}$ and $\ddot{W}_{inm}$ terms.  In general there are
counterterms with more complex structure, involving higher order Bessel
functions, but they are largely degenerate on the scales where our 1-loop
calculation is applicable (see also \cite{VlaWhiAvi15}) so we include only
$\xi_L$ above.  We find these terms to be small, numerically.
The predictions for $\sigma^2_{ij}$ are weakly dependent on the bias,
except the dependence that comes through $\alpha_\sigma$.

In Refs.~\cite{ReiWhi11,WanReiWhi14} it was noted that perturbation theory
compares relatively well with numerical simulations on the shape of the
pairwise velocity dispersion, but does much less well on the overall
amplitude.  The difference between the N-body measurements and the
predictions of both Eulerian and Lagrangian perturbation theory is close
to a constant.
Here we see that the leading order EFT term, $\alpha_\sigma\delta_{ij}$,
changes
\begin{equation}
  \sigma^2_{ij} \to \sigma^2_{ij} + \alpha
  \ \frac{1+\xi_{\rm mat}^{\rm 0-loop}}{1+\xi_{\rm halo}^{\rm 1-loop}}
  \ \delta_{ij} \quad .
\end{equation}
On large scales ($\xi\ll 1$) this is equivalent to adding a constant to both
$\sigma^2_\parallel$ and $\sigma^2_\perp$ and drastically improves the
agreement between theory and simulation\footnote{In principle this counter
term can be negative and lead to $\sigma_{12}^2<0$ at small scales.  We have
not found this to be an issue in practice in our case.  Should this become
an issue one can keep just the linear part of $\sigma_{12}^2$, including the
positive $\alpha_\sigma$ contribution, in the exponent and determinant and
expand in the higher orders.  This guarantees positive definiteness and
introduces corrections only at higher orders higher than we are keeping.}.
This term corresponds to the lowest order correction for the finger of
god effect suggested by Ref.~\cite{ReiWhi11}.
Fingers of god are one of several small-scale effects missing from the
perturbative treatment that can be handled using EFT methods.
As noted above the $\alpha_\sigma$ counterterm has (zero lag)
contributions from multiple terms:
$ \ddot{A}_{nm}$, $\ddot{A}^{10}_{nm}$ and $\ddot{\Upsilon}_{mn}$.
Thus it can have a non-trivial bias dependence.

The $\alpha_\sigma$ term appears as a $\delta$-function in Fourier space, as
does the $\alpha_\sigma P_{\rm Zel}$ contribution
(which corresponds to the leading $P_{02}$ contributions in \cite{Vla12,Vla2013},
see also Appendix~\ref{app:fourier}).
When considering $M_{0}$ and $M_{1,n}$, we have been assuming that the
$b_{\nabla^2}$ terms whose leading order contributions are $k^2\,P_L$ are
of similar size to the one-loop terms of similar form.
In $M_{2,nm}$ these $b_{\nabla^2}$ terms enter multiplied by terms like
$U^{(1)}$ and hence can be dropped.
This, in principle, breaks the degeneracy of the EFT counterterms and 
these derivative bias terms.

\subsection{Impact of parameters}

In this section we take a closer look at how the various parameters
affect the predictions of the ingredients going into the GSM.

\begin{figure}
\begin{center}
\resizebox{\columnwidth}{!}{\includegraphics{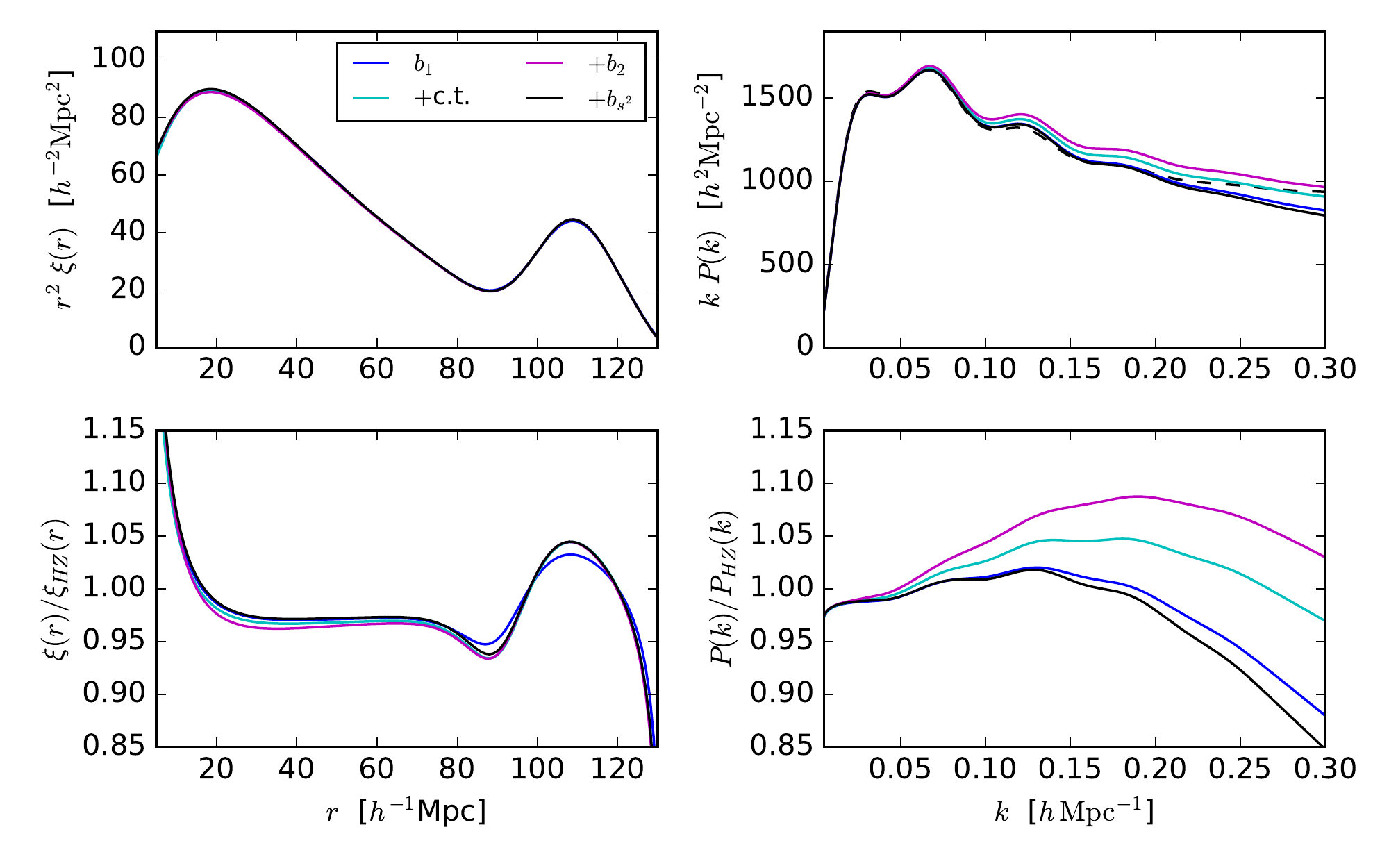}}
\end{center}
\caption{The contribution of different terms to the real-space 2-point
function in (left) configuration and (right) Fourier space.  The upper
panels show $r^2\,\xi(r)$ or $k\,P(k)$ while the lower panels show the
ratio to a fiducial theory (taken to be the Halo-Zeldovich model, the
black dashed line in the upper right panel; see text).
The blue lines show the predictions with $b_1=1$ and all other parameters
set to zero.  Including the counter-term with $\alpha=-20$ (to emphasize the
effect) generates the cyan line. Further adding $b_2=0.25$ gives the magenta
line and finally adding $b_{s^2}=0.5$ gives the black line.}
\label{fig:real}
\end{figure}

One of the earliest models of structure formation is the Zeldovich
approximation, which does a good job of describing the growth of large
scale structure into the non-linear regime (see e.g.~Ref.~\cite{Whi14}).
Our theoretical model has several terms beyond the Zeldovich approximation,
including 1-loop corrections in LPT, corrections for missing small-scale
physics and a generalized, non-linear bias parameterization.
In this subsection we will turn on these terms one at a time so that their
impact on the clustering predictions can be assessed (see also
Fig.~\ref{fig:xir_pieces}).

Fig.~\ref{fig:real} compares models for the real-space clustering in
both Fourier (see Appendix \ref{app:fourier}) and configuration space.
To better show small deviations from the lowest order result we also show
residuals compared to a reference model, and we take as the reference to
be the Halo-Zeldovich model \cite{SelVla15} in the form
\begin{equation}
  P_{HZ}^{hh}(k) \equiv (1+b_1)^2 P_{\rm Zel}^{m}(k) + {\rm const}
\label{eqn:pkhz}
\end{equation}
where $P_{\rm Zel}^m$ is the Zeldovich approximation for the mass power
spectrum \cite{BonCou88,FisNus96,TayHam96,CLPT,Tas14,SelVla15}.
This model takes out the large-scale trends, allowing us to focus on the
features in more detail.
Notice that the const term in $P(k)$ transforms to a $\delta$-function at
zero lag in the correlation function, so our reference model for $\xi(r)$
is simply the Zeldovich matter correlation function times $(1+b_1)^2$.

The blue line shows the effect of including lowest order Lagrangian bias
($b_1$) compared to simply multiplying the matter result by $(1+b_1)^2$.
We then add the EFT counter term (cyan line), which is degenerate with the
$b_{\nabla^2}$ terms, which affects the width of the peak in configuration
space and adds additional power at higher $k$ with undamped BAO oscillations
in Fourier space.  Both of these effects are expected from the nature of the
term, $k^2\,P_L(k)$ or a Laplacian of $\xi_L(r)$.
We have chosen a very large value of $\alpha$ in order to emphasize the
effect, and the sharpening of the peak is quite pronounced in the figure.
Adding $b_2$ has little effect at the BAO peak but reduces the correlation
function at smaller $r$.  Including $b_2$ adds significant power to $P(k)$
at high $k$.
Finally we add $b_{s^2}$.  The dominant effect in configuration space is a
steepening of the correlation function, with a small change in the shape of
the BAO peak.  In Fourier space the shear terms modify high $k$ significantly
like the $b_2$ terms do.

We don't show the impact of the parameters on $v_{12}$ and $\sigma_{ij}^2$
in a figure, as very few terms matter.
On large scales $v_{12}$ is proportional to $1+b_1$, with the $b_2$ terms
having very little effect.  Either of the two counter terms causes a
steepening of $v_{12}$ at small $r$, which we will see improves agreement
with the N-body and is important for matching the quadrupole to small scales
(see discussion in Ref.~\cite{ReiWhi11}).
The parallel and transverse components of the dispersion are very
insensitive to any of the bias terms.
The EFT counter term amounts to adding (or subtracting) a constant to both
$\sigma_\parallel^2$ and $\sigma_\perp^2$, which helps improve agreement
with the simulations and matches the finger-of-god correction implemented
by Ref.~\cite{Rei12}.

\subsection{The bias model and advantages of a Lagrangian formulation} 
\label{sec:bias_philosophy}

We end this section with a few words about the bias model.
When comparing to observations of biased tracers of the density field,
choosing a suitably powerful and flexible bias model is clearly essential
in order to not introduce biases into a fit.  We have seen above that the
relative size of even quite high-order bias terms is comparable to the
affects of gravitational non-linearity at 1-loop or beyond.

The most general approach to bias is an EFT-inspired one
(see \cite{McDRoy09,Ass14,Sen14,Mirb14,Ang15}).
In fact, reasoning purely from symmetry is even more attractive for the
bias model than for perturbative non-linearity because the ultimate theory
we are wishing to describe in the case of bias is considerably more complex
than the N-body problem.  However, these approaches have the huge drawback
of an explosion of undetermined parameters, especially as we push to higher
order and include terms like $\nabla^2\delta$ and $s_{ij}$ in our expansion.
When the fitting range and accuracy are constrained, and only a small number 
of statistics are fit for, the difficulties of exploring multidimensional
likelihood surfaces with strong degeneracies can become extreme.

While some gains can be made by including additional statistics (e.g.~the
3-point functions) in the fits, we believe the appropriate approach is a
combination of symmetry-based parameterizations with priors based on
physical models of bias
(e.g.~Refs.~\cite{Kai84,BBKS,ColKai89,MoWhite96,ST99}).
By allowing a flexible bias scheme, one can explore evidence from the data
of violations of the simplest models.  By including priors based on our
decades of exploration of biasing in simulations and observations we can
mitigate (to some extent) the difficulties inherent in any high-dimensional
scheme.

In this approach a Lagrangian formulation of the problem appears to us to
be highly beneficial.  In cosmology the initial conditions are often
considerably less complex than the late-time observations one is trying
to describe.  By formulating our bias model in terms of the Lagrangian
field, and by disentangling the effects of evolution from complex bias,
the Lagrangian formulation allows a more straightforward implementation of
the programme described above.

\section{Evolution and the effective-redshift approximation}
\label{sec:evolution}

Observations unavoidably measure clustering not at fixed time but
along the past light-cone.  Since the structure, and the bias of objects,
evolves with time the measured correlation function is an ``effective'' one.
If we define
\begin{equation}
  X_{\rm eff} = 
  \frac{\int dz(dN/dz)^2(H/\chi^2)\ X(z)}
       {\int dz(dN/dz)^2(H/\chi^2)}
\label{eq:Xeff}
\end{equation}
then on scales small compared to the scale of variation of $dN/dz$ we measure
$\xi_{\ell,{\rm eff}}$ \cite{Mat97,WhiMarCoh08,Whi15b} which we interpret as
$\xi_\ell(z_{\rm eff})$, depending on parameters $\theta_{\rm eff}$.
The accuracy of this approximation depends on the width of $dN/dz$ and
how rapidly the bias parameters (and the sample) are changing.


We use our analytic theory to investigate this issue, taking flat $dN/dz$
of width $\Delta z=0.1$ and 0.2.
We have found that, for smoothly varying bias parameters, approximating
$\xi_{\ell,{\rm eff}}$ as $\xi_\ell(\theta_{\rm eff},z_{\rm eff})$ induces
very small errors, typically below one percent even for $\Delta z=0.2$.
This is true whether we take $\theta(z)$ to be constant, or impose an
evolution based upon the peak-background split and fixed peak height
\cite{Kai84,ST99} or use the theory of excursion sets
\cite{Kai84,ColKai89,MoWhite96,ST99} or Gaussian peaks \cite{BBKS}.
In making this comparison we need to assume a redshift-dependence of the
counter terms, $\alpha_i$.  This is in principle arbitrary.
While we know that the piece which absorbs the cut-off dependence must scale
as $D^2(z)$, any finite pieces could evolve in a different way.
Morevover, our $\alpha_i$ also account for $\nabla^2$-bias terms, which are
not expected to be proportional to $D^2(z)$. 
Despite these complications, we assume $\alpha_i\propto D^2(z)$, which is
close to what is measured in N-body simulations \cite{Baldauf15}.

For these assumptions, within our model, approximating $\xi_{\ell,{\rm eff}}$
by $\xi_\ell(\theta_{\rm eff},z_{\rm eff})$ induces sub-percent errors for
both the monopole and the quadrupole.
This suggests that, for such smoothly evolving parameters, the neglect of
light-cone effects is subdominant to higher order terms in the perturbative
expansion or bias expansion.  Of course, the redshift evolution of a real
galaxy sample could be much more complicated than the models we have used,
since several observational effects could start to play a role.  Such
situations would need to be investigated on a case-by-case basis, but the
formalism above allows a rapid exploration of these approximations.

\section{Comparison with simulations}
\label{sec:nbody}

We now make a preliminary comparison of the performance of the GSM,
and its components predicted in perturbation theory, to the clustering
of halos from N-body simulations.
We find that we are currently limited by the size of systematic errors in
the simulations, and so we defer a more detailed comparison to future work.

We make use of the halo catalogs\footnote{The data are available at
http://www.hep.anl.gov/cosmology/mock.html.  Of the 5 realizations, the
data for the first were corrupted so we used only the last 4.} from the
simulations described in Ref.~\cite{Sun16}.
Briefly, there were 4 realizations of a $\Lambda$CDM
($\Omega_m=0.2648$, $\Omega_bh^2=0.02258$, $h=0.71$, $n_s=0.963$,
 $\sigma_8=0.8$)
cosmology simulated with $4096^3$ particles in a $4\,h^{-1}$Gpc box.
The simulations were run with ``derated'' time steps chosen so that the
matter power spectrum is accurate to better than 1\% out to
$k=1\,h\,{\rm Mpc}^{-1}$ and halo masses were adjusted to match the
halo abundance of a simulation with finer time steps.

We retrieved the halo catalogs for $z=0.8$ and $0.55$.
The halos were defined by a friends-of-friends algorithm \cite{DEFW} with a
linking length of $0.168$ times the mean interparticle spacing (and the
masses were redefined, as described in \cite{Sun16}).

Ref.~\cite{Sun16} did not present convergence tests for the multipoles of
the halo correlation function or power spectrum (though their Fig.~6 compares
the monopole of the cross-correlation of the real- or redshift-space halo
field with the real-space matter field).
However the authors kindly made available two sets of simulations, one with
derated time steps and one with time steps they consider ``converged''.
By comparing the clustering of halos, with remapped masses, in the two
simulations we can estimate\footnote{Our estimate is uncertain because the
two simulations were evolved from different initial conditions, and thus
some of the differences between the runs are due to sample variance and not
methodological differences.  However, the very large size of the boxes
mitigates this uncertainty to some extent.} the effects of the derated time
steps and hence a systematic error.
The real-space power spectrum and the monopole of the redshift-space power
spectrum show coherent, oscillatory residuals at the 1-2\% level with
slowly decreasing amplitude to $k\simeq 0.5\,h{\rm Mpc}^{-1}$.
The power spectrum quadrupole shows oscillatory signals at just over 2\%,
with a feature near the acoustic scale.
The effects are much larger in configuration space, with the ratio of the
multipole moments differing from unity by 5\% over much of the range of
interest to us, with differences of more than 10\% near $80\,h^{-1}$Mpc.
The impact of these artifacts is likely highly dependent upon the use to
which the simulations are being put, and quite difficult to assess  without
doing a detailed fit to a specific template.  In our case, where we want to
make precise predictions at large scales, they limit us to semi-qualitative
comparisons below.

It is worthwhile to note that the total volume simulated,
$256\,h^{-3}{\rm Gpc}^3$, is equivalent to $>40$ and $>25$ full-sky
surveys for redshift slices $0.5<z<0.6$ and $0.75<z<0.85$, respectively,
and approximately $50\times$ the total volume of the BOSS survey \cite{Daw13}.
The statistical errors from the simulations should thus be much smaller
than those of any future survey confined to a narrow redshift slice
(see \S\ref{sec:evolution}),
and are dominated by systematic errors in the algorithms or physics
missing from the simulations themselves.

We first computed the real- and redshift-space power spectrum and correlation
function multipoles for halos in bins of mass
($12.5<{\rm lg} M<13.0$, $13.0<{\rm lg}M<13.5$ and $13.5<{\rm lg}M<14.0$;
 all masses in $h^{-1}M_\odot$)
using the full periodic box, and used the average and scatter from the 4
boxes as our estimate of the signal and statistical uncertainty.
We compute the power spectra in bins of width $0.0031\,h\,{\rm Mpc}^{-1}$
and the correlation functions in bins of width $2\,h^{-1}$Mpc, which
is small enough that effects due to binning are $\mathcal{O}(0.1\%)$ for the
theories we wish to test.
For the best measured, lowest mass, sample the statistical uncertainty on
the redshift-space correlation function is at most a few percent on the
scales of interest
($<1\%$ for $s<50\,h^{-1}$Mpc; $\sim 3\%$ at $s\simeq 80\,h^{-1}$Mpc;
and $\sim 2-3\%$ for $s>100\,h^{-1}$Mpc for the redshift-space monopole for
the lowest mass sample with smaller fractional errors for the quadrupole),
with larger errors for the rarer halo samples (growing to tens of percent at
high-$s$ for the rarest samples).
For the real-space correlation function the errors on the best-measured
sample are $<1\%$ for $r<60\,h^{-1}$Mpc and still $<2\%$ across the acoustic
peak.  These rise to 4\% across the acoustic peak for the highest mass sample.
Similarly we computed the mean pairwise velocity and dispersion as a function
of separation (in 100 bins equally spaced in $r$) for halos in the same mass
bins.
The mean pairwise velocity and dispersion were determined, statistically, to
better than a percent on all relevant scales (though the systematic error from
the derated time steps and halo mass is larger).
The power spectrum used to generate the simulation intial conditions was not
available to us, so we used CAMB\footnote{http://cosmologist.info/camb}
\cite{CAMB} to compute the linear theory power spectrum (using the
cosmological parameters above) which is needed for the perturbation theory
predictions.

We first checked that the lowest order streaming model is accurate when
the ``true'' ingredients from the simulation are used.
This updates the tests done in \cite{ReiWhi11,Uhl15}.
Specifically we took the $\xi$, $v_{12}$ and $\sigma_{12}$ measured from
the simulations and used the GSM as in Eq.~(\ref{eqn:gsm}) to generate the
multipoles of the correlation function.  As had been found previously, the
GSM works well, lying within $\pm 2\%$ of the simulations on scales above
$25\,h^{-1}$Mpc for all masses and redshifts shown (i.e.~within the systematic
error budget of the simulations themselves).
The monopole performs exceptionally well all the way down to $20\,h^{-1}$Mpc.
Below this scale the model requires extrapolating $v_{12}(r)$ and
$\sigma_i^2(r)$ to very small scales where it is hard to measure directly.

\begin{figure}
\begin{center}
\resizebox{\columnwidth}{!}{\includegraphics{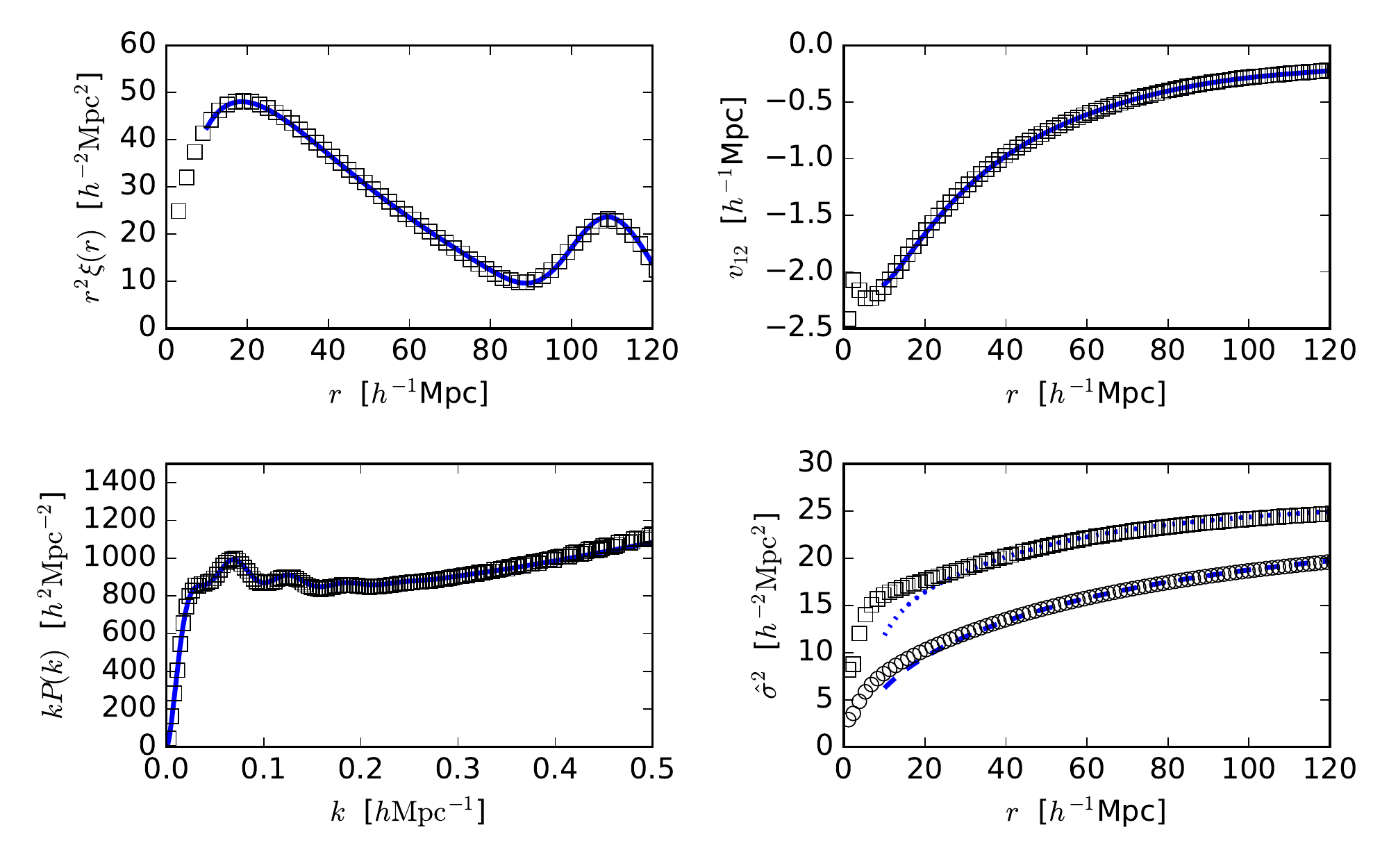}}
\end{center}
\caption{A comparison of the ingredients of the GSM, as computed in LEFT,
with N-body simulations.  We have chosen $z=0.55$ and $12.5<{\rm lg}M<13.0$
as it has the smallest errors in the N-body (the other redshifts and mass
ranges are qualitatively similar).
(Top Left) The real-space correlation function, (Top Right) the mean infall
velocity, (Bottom Left) the real-space power spectrum and (Bottom Right)
the velocity dispersions parallel to (dotted) and transverse to (dashed)
the separation vector.  In each case the lines show the analytic prediction,
while the points show the N-body results.
The agreement on large scales is good, comparable to the numerical
accuracy of the N-body results themselves.}
\label{fig:cmp_nbody}
\end{figure}

Now we check how well perturbation theory predicts each of the ingredients
going into the streaming model.
This updates the tests done in \cite{ReiWhi11,WanReiWhi14,Uhl15}.
Overall the agreement between the theory and simulations is very good.
We find this for all redshifts and mass bins that we have checked.
Fig.~\ref{fig:cmp_nbody} compares the model predictions for the real-space
halo correlation function and power spectrum, the mean infall velocity and
the velocity dispersions to the N-body results at $z=0.55$ for halos with
$12.5<{\rm lg}M<13.0$.  We show this sample as it has the smallest statistical
errors in the simulations, but the results for the other redshifts and mass
sample are very similar.  Errors on the N-body points are omitted for clarity.
The statistical errors are approximately the size of the points.  As we have
argued above, the systematics errors are at the several per cent level.

Fig.~\ref{fig:cmp_nbody} also compares the N-body and model results
for the mean pairwise velocity, and again the agreement is very good.
We also find that the model captures the scaling with mass and redshift well.
Inclusion of the $\nabla\xi$ counter term improves the agreement with the
N-body results on small scales, steepening the slope of $v_{12}$ at small $r$.
As has been noted before \cite{ReiWhi11}, correctly predicting the slope of
$v_{12}$ is crucial to modeling $\xi_2$ well in the GSM.

Finally Fig.~\ref{fig:cmp_nbody} also compares the N-body and model
predictions for the pairwise velocity dispersion ($\hat{\sigma}^2_i$)
parallel and transverse to the separation vector.
As noted in Ref.~\cite{ReiWhi11}, the theory without the EFT terms
mistestimates the dispersion.  However the misestimate is very close to a
constant, as expected from the lowest order EFT correction.
This EFT correction is degenerate with the finger-of-god correction, as
implemented by Ref.~\cite{Rei12}, which is also a non-linear effect.  As
such it does not increase the number of parameters in the model.

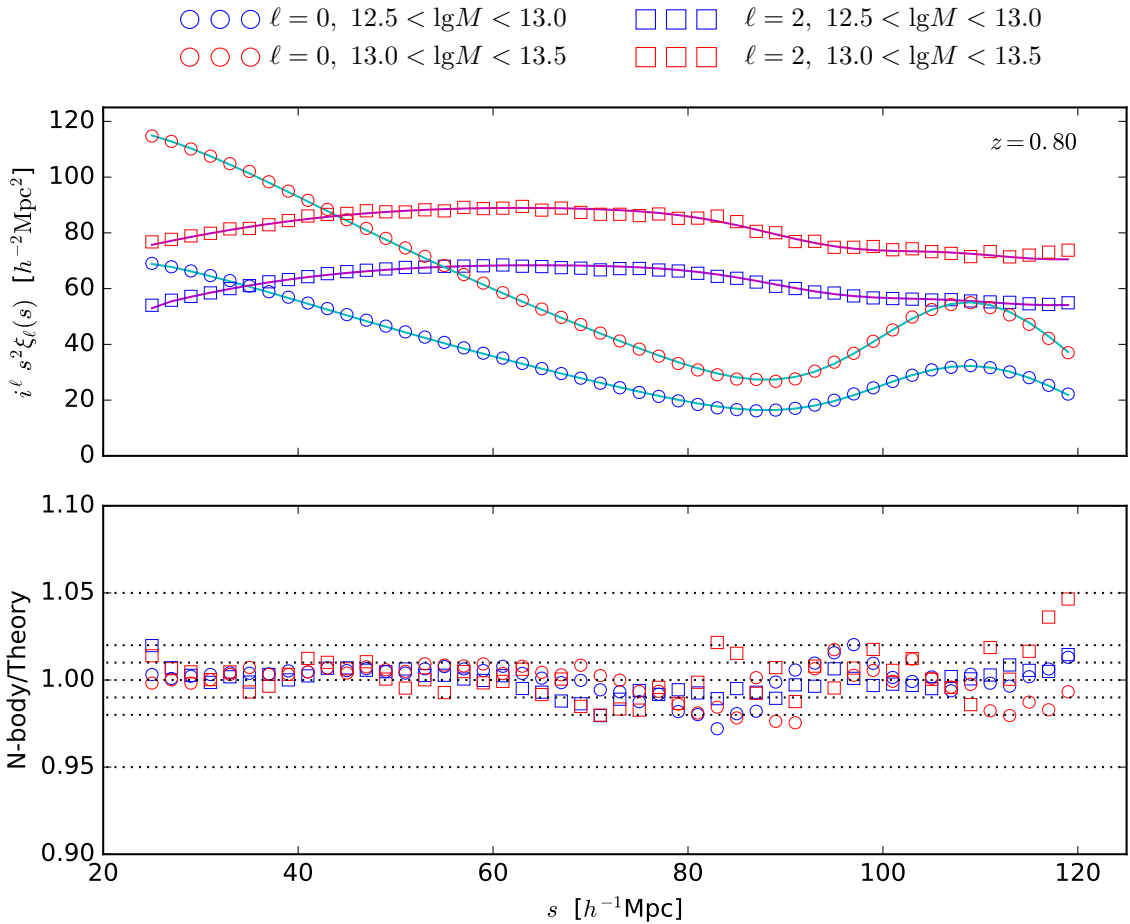
\begin{figure}
\begin{center}
\hspace{1.2cm}
\begin{tikzpicture}
\draw[red] (0,0) circle (0.12cm);
\draw[red] (0.4,0) circle (0.12cm);
\draw[red] (0.8,0) circle (0.12cm) node[black, right] {\small $~\ell = 0,~13.0<{\rm lg}M<13.5$} ;
\draw[blue] (0,0.5) circle (0.12cm);
\draw[blue] (0.4,0.5) circle (0.12cm);
\draw[blue] (0.8,0.5) circle (0.12cm) node[black, right] {\small $~\ell = 0,~12.5<{\rm lg}M<13.0$} ;
\node (rect) at (6,0) [draw,red]{};
\node (rect) at (6.4,0) [draw,red]{};
\node (rect) at (6.8,0) [draw,red]{}; \node at (9.2,0) {\small $~\ell = 2,~13.0<{\rm lg}M<13.5$};
\node (rect) at (6,0.5) [draw,blue]{};
\node (rect) at (6.4,0.5) [draw,blue]{};
\node (rect) at (6.8,0.5) [draw,blue]{}; \node at (9.2,0.5) {\small $~\ell = 2,~12.5<{\rm lg}M<13.0$};
\end{tikzpicture}\\
\resizebox{\columnwidth}{!}{\includegraphics{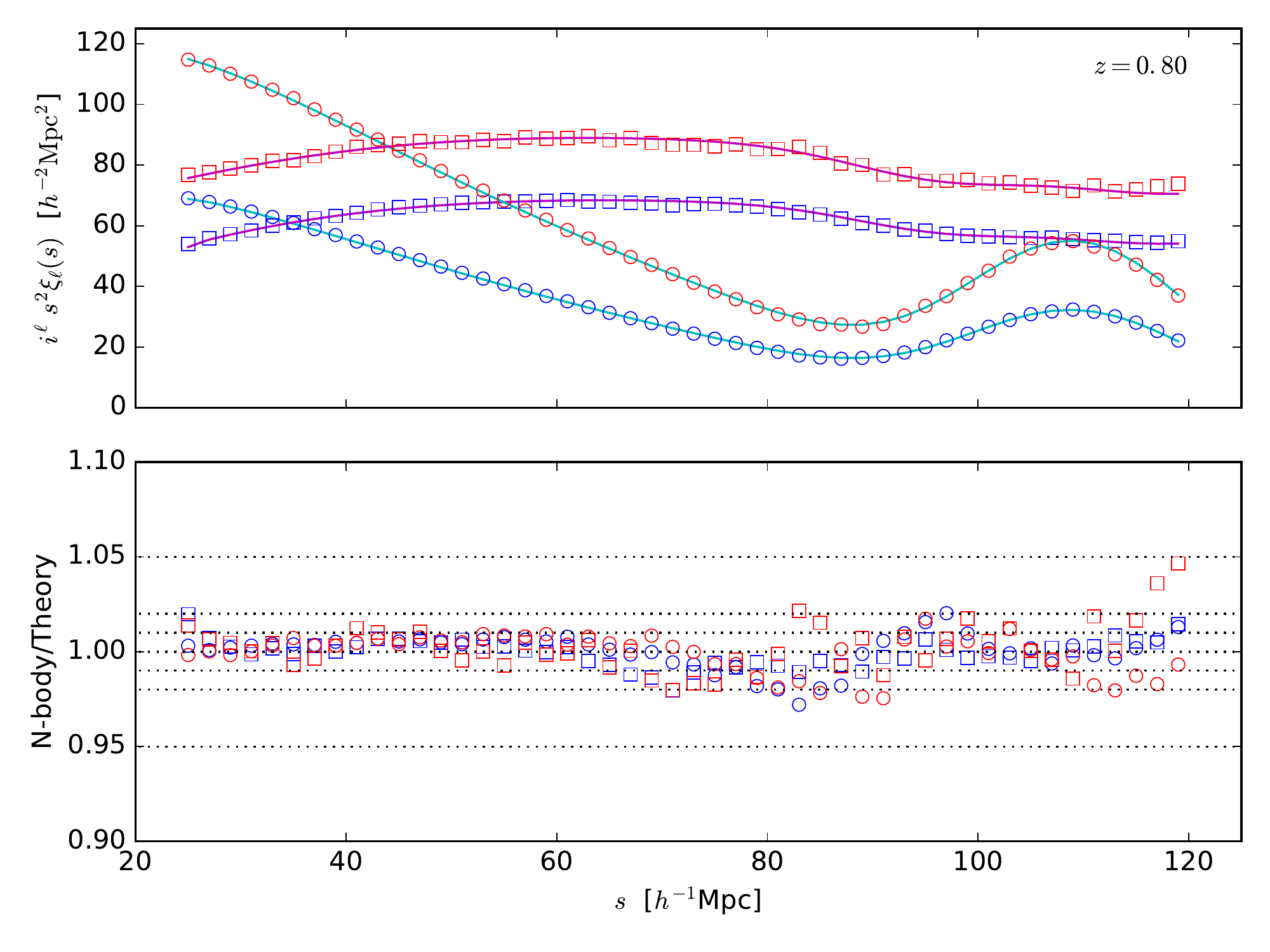}}
\end{center}
\caption{The comparison of the multipole moments of the redshift-space
correlation function, $\xi_\ell(s)$, with the N-body results.
The circles indicate the monople moment, the squares the quadrupole moment
while blue points are for the $12.5<{\rm lg}M<13.0$ mass bin and red points
the $13.0<{\rm lg}M<13.5$ bin.  The fit is only to the points above
$30\,h^{-1}$Mpc, but the model clearly provides a good fit to the data below
this scale as well.}
\label{fig:xi_ell}
\end{figure}

\begin{table}
\begin{center}
\begin{tabular}{c|cccccc}
Mass & $b_1$ & $b_2$ & $b_{s^2}$ & $\alpha_\xi$ & $\alpha_v$ & $\alpha_\sigma$
\\ \hline
$12.5<{\rm lgM}<13.0$ & 0.68 & -1.01 & -0.92 & -24 & -52 & -18 \\
$13.0<{\rm lgM}<13.5$ & 1.28 & -1.34 & -0.14 &  -9 &  25 &  -3
\end{tabular}
\end{center}
\caption{The best-fit parameters of the model shown in Fig.~\ref{fig:xi_ell}.
These parameters have been fit `by eye' and thus should be taken as
indicative.}
\label{tab:fit_params}
\end{table}

Finally we test the whole model.  Fig.~\ref{fig:xi_ell} shows the monopole
and quadrupole of the redshift-space correlation function for two mass bins
at $z=0.8$ (the parameters are given in Table \ref{tab:fit_params}).
This is the redshift bin for which we had the worst agreement with the
ingredients, so gives the most pessimistic view of the model performance.
We show the absolute agreement in the upper panel and the ratio of the
N-body to the theory in the lower panel so that small deviations can be seen.
Note in all cases the level of agreement is comfortably within the systematic
errors of the simulations, which we estimate to be several per cent (we have
shown $\pm 1$, $\pm 2$ and $\pm 5\%$ as dotted lines to guide the eye).

\section{Conclusions}
\label{sec:conclusions}

Redshift surveys by neccessity measure large-scale structure in redshift
space, in which peculiar velocities sourced by large-scale gravitational
potentials have induced anisotropic clustering.
Measurement of these anisotropies allows us to probe the growth of
large-scale structure, breaking degeneracies in cosmological distance
measures and providing a key test of general relativity and the gravitational
instability paradigm on quasi-linear, cosmological scales.

In this paper we update earlier treatments of the Gaussian Streaming Model
(GSM) and present tests against a new set of N-body simulations.
We improve upon previous calculations of the ingredients in this model --
the real-space correlation function, mean pairwise velocity and pairwise
velocity dispersion -- using a Lagrangian effective field theory with an
extended bias model.
We show that the Lagrangian approach provides a solid framework for studying
large-scale structure, and provides a simple connection to N-body simulations
and peaks theory.  Effective field theory techniques provide a straightforward
means of incorporating the effects of non-perturbative physics into
perturbation theory by including additional terms whose structure is
determined by the symmetries of the theory.
The expressions for the ingredients, and the bias model, are new and
present the most general expressions at the given order\footnote{Code to
evaluate these expressions is available at
{\tt https://github.com/martinjameswhite/CLEFT\_GSM}.}.

Throughout our focus has been on increasing the precision with which we
can predict the clustering moments on intermediate scales ($>25\,h^{-1}$Mpc),
rather than on increasing the range of scales we predict.
We believe this is the most appropriate use of techniques built upon
perturbation theory.  Ultimately the precision of our model is limited by
the neglect of 2-loop terms in the perturbative calculation, higher
derivative orders in the EFT expansion, higher order terms in the bias
expansion and the neglect of lightcone evolution.  We find that these
effects alter the predictions for the monopole and quadrupole moment of the
correlation function and power spectrum at the per cent level on scales
above $25-30\,h^{-1}$Mpc.

The inclusion of 1-loop corrections to the Zeldovich approximation changes
the clustering statistics by several per cent on large scales.
The EFT terms encapsulate the effects of small-scale physics which is
missing from the standard perturbative treatment.  In $\xi(r)$ the primary
effect is to change the width of the BAO peak and slightly decrease $\xi$
at lower $r$.  The EFT terms steepen $v_{12}$ at small $r$, which is important
in the streaming model in order to match the quadrupole.
The EFT terms are most important for $\sigma^2$, where there is a large
mismatch between the perturbative prediction and the N-body results
\cite{ReiWhi11,WanReiWhi14}.  The difference is very close to a constant,
independent of scale and orientation, which is also the behavior of the
lowest-order EFT counter term.  Such an offset was included in earlier
versions of the GSM as a ``finger-of-god'' term, referring to a specific
type of small-scale effect.

We find that a flexible bias model is at least as important as including
the higher-order contributions to the evolution of clustering.
The most general approach to bias is an EFT-inspired one, and reasoning
purely from symmetry is highly attractive when describing the complex physics
which leads to bias.
We use a Lagrangian bias expansion up to the second order, including
a derivative term and a shear term. 
This generates all the terms present in the corresponding third order
Eulerian biasing expansion (see e.g.~Refs.~\cite{McDRoy09,Sen14,Mirb14,Ang15}),
although the latter has more freedom coming from an additional third order
bias parameter in the real space two-point statistics. 
Adding the third order terms in Lagrangian space would yield the same number
of free parameters in both Lagrangian and Eulerian picture.
We argue that using a symmetry-based approach to bias, with priors set by
theory and simulations, has many benefits.  In such a scheme, a Lagrangian
framework has multiple advantages over an Eulerian one
(\S\ref{sec:bias_philosophy}).

We have compared our theoretical calculations with 4 large N-body simulations
\cite{Sun16}, with a total simulated volume of $256\,h^{-3}{\rm Gpc}^3$.
This volume, many times larger than accessible observationally, leads to
very small statistical errors.  However the simulations were run with an
approximate time-stepping scheme, which limits the overall accuracy to
several percent on the statistics (and scales) of relevance here.
With this caveat in mind, our model performs very well when compared to
N-body simulations.

The model presented here achieves per cent level accuracy on the monopole
and quadrupole of the correlation function and power spectrum on quasi-linear
scales.
This level of accuracy is likely sufficient for all upcoming surveys --
going to higher order in perturbation theory or including additional EFT
terms will yield little return.
In order to push to smaller scales, detailed modeling of highly-nonlinear
effects are required (e.g.~Refs.~\cite{Rei14,Oku15})
which will likely increase the number of parameters dramatically for even
a small increase in dynamic range.
Increasing the volume, to decrease the errors at fixed scale, requires
increasing the redshift range and requires modeling of the evolution of the
bias (in addition to survey non-idealities).

Our model works in a Lagrangian framework, with parameters that are easy
to interpret within the context of the Gaussian peaks formalism or N-body
simulations.  Along with the LPT-based model for post-reconstruction BAO
presented in \cite{Whi15a}, this formalism can be used to interpret the
measurements from upcoming redshift surveys.
The analytic nature of the calculation makes it possible rapidly explore
changes in cosmology, and the flexible, parameterized bias model allows
exploration of a wide range of effects with little effort.  The analytic
calculation can be used to set requirements for future grids of N-body models,
both in terms of modeling the response surface for an emulator and in terms
of which modes and which statistics need to be well converged.

\acknowledgments
We would like to thank Ravi Sheth and Uros Seljak for useful discussions
during the preparation of this manuscript.

Z.V.~is supported in part by the U.S. Department of Energy contract to
SLAC no.~DE-AC02-76SF00515.

This research has made use of NASA's Astrophysics Data System.
The analysis in this paper made use of the computing resources of the
National Energy Research Scientific Computing Center.

\appendix

\section{The Gaussian Streaming Model}
\label{app:GSM}

\begin{figure}
\begin{center}
\resizebox{\textwidth}{!}{\includegraphics{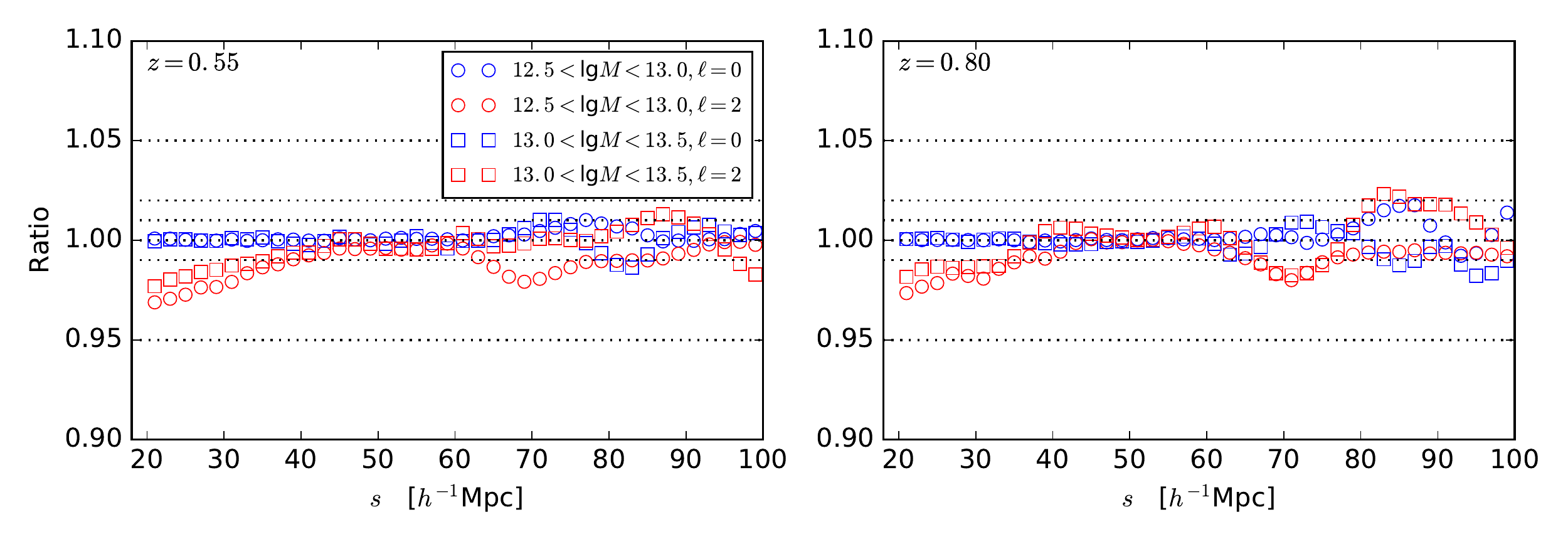}}
\end{center}
\caption{A test of the GSM using $\xi(r)$, $v_{12}(r)$ and $\sigma^2_{ij}(r)$
measured directly from the simulation.  We have measured the ingredients of
the GSM from the halo catalogs of the four N-body simulations, averaged them
and then smoothed them to reduce noise.  These were then used as ingredients
in the GSM to compute $\xi_\ell(s)$.  The figure shows the ratio of the
$\xi_\ell(s)$ measured directly in the simulations (and smoothed to reduce
noise) to the GSM prediction.
The agreement on the monopole is excellent.  The agreement on the quadrupole
is within $\pm 2\%$ above $25\,h^{-1}$Mpc.}
\label{fig:test_gsm}
\end{figure}

There are several routes to deriving the GSM, and its generalizations.  The
continuity equation allows us to relate the 2-point function in redshift
space to a generating function \cite{Fis95,ReiWhi11,Uhl15}.  If we recall
that a shift in configuration space generates a phase in Fourier space, the
generating function is
\begin{equation}
  1+\mathcal{M}(\mathbf{J},\mathbf{r}) =
  \left\langle\left[1+\delta(\mathbf{x})\right]
  \left[1+\delta(\mathbf{x}')\right]
  e^{i\mathbf{J}\cdot\Delta\mathbf{u}}\right\rangle
\end{equation}
where $\mathbf{r}=\mathbf{x}-\mathbf{x}'$,
$\Delta\mathbf{u}=\mathbf{u}(\mathbf{x}')-\mathbf{u}(\mathbf{x})$,
$\mathbf{u}$ is the velocity field in units of the Hubble expansion and
$\mathbf{J}=k_\parallel\hat{z}$.
Fourier transforming $1+\mathcal{M}(k_\parallel\hat{z},\mathbf{r})$
gives the redshift-space power spectrum.
Using the cumulant expansion, i.e.~expanding
$\ln\left[1+\mathcal{M}\right]$ in powers of $\mathbf{J}$, we have
\begin{equation}
  1+\mathcal{M} = \exp\left[\sum_{n=0}^\infty
  \frac{i^n}{n!}k_{i_1}\cdots k_{i_n} \mathcal{C}^{i_1\cdots i_n}\right]
\end{equation}
where $C^{i_1\cdots i_n}$ are the density-weighted velocity cumulants.  The
first few cumulants are
\begin{eqnarray}
  C      &=& \ln\left[1+\xi\right] \\
  C^i    &=& \frac{\langle(1+\delta)(1+\delta')\Delta u^i\rangle}{1+\xi}
          \equiv v_{12}^i \\
  C^{ij} &=& \frac{\langle(1+\delta)(1+\delta')\Delta u^i\Delta u^j\rangle}
                  {1+\xi}
          - v_{12}^i v_{12}^j \equiv  \sigma_{12}^{ij}
  \label{eq:cums}
\end{eqnarray}
Keeping only terms through $n=2$ in $1+\mathcal{M}$, and performing the
Fourier transform, the redshift-space correlation function is
\begin{equation}
  1+\xi(\mathbf{s}) = \int d^3r\left[1+\xi(\mathbf{r})\right]
  \int\frac{d^3k}{(2\pi)^3}
  \ e^{-ik_i(s_i-r_i-v_{12}\hat{z}_i)}
  \ e^{-(1/2)k_ik_j\sigma_{12}\hat{z}_i\hat{z}_j}
  \label{eq:gsm}
\end{equation}
The $d^3k$ integral can be performed analytically and the result is the GSM,
given in the main text.  Higher order cumulants introduce higher corrections.
These higher cumulants could be significant in case of large dispersion
objects, and low redshifts.
For example dark matter particles exhibit large FoG effects and some of
these would be decomposed into these higher cumulants. 
Similar effects, but to smaller extent, might show up in smaller size halos,
while we expect the higher terms to be less relevant for more massive halos 
and higher redshifts.
The performance of the GSM when $\xi(r)$, $v_{12}(r)$ and $\sigma^2_{ij}(r)$
are measured from the simulations are shown in Fig.~\ref{fig:test_gsm}.

\section{Fourier space representation}
\label{app:fourier}

The text focuses on the configuration-space description of the model and
the clustering statistics, however on numerous occasions we have found it
useful to examine the Fourier-space versions as well.  In this appendix we
give an example of the relevant formulae, referring the reader to
Refs.~\cite{CLPT,WanReiWhi14,VlaWhiAvi15,VlaSelBal15} for further details.

For example, the real-space, power spectrum (for $k\ne 0$) of halos can be written
\begin{align}
P^{hh}(k) &= \int d^3q\ e^{i\mathbf{k}\cdot\mathbf{q}}
  \exp\left[-\dfrac{1}{2}k_ik_j A_{ij}^{\rm lin}\right] \Bigg\{
  1 - \frac{1}{2} k_ik_j A^{\rm 1-loop}_{ij} - \frac{i}{6} k_ik_jk_l W^{\rm 1-loop}_{ijk} \non\\
&~~~
- b_{1} \Big( k_i k_j A^{10}_{ij} - 2 i k_i U^{(1)}_i \Big) 
+b_{1}^2 \Big( \xi_L + i k_i U^{11}_i - k_i k_j U^{(1)}_i U^{(1)}_j  \Big) \non\\
&~~~ 
+ b_{2} \Big( i k_i U^{20}_i - k_i k_j U^{(1)}_i U^{(1)}_j  \Big)
+ b_{1} b_{2} \Big( 2 i k_i U^{(1)}_i  \xi_L \Big) 
+ b_{2}^2 \bigg( \frac{1}{2} \xi_L^2 \bigg) \non\\
&~~~ 
- b_{s^2} \Big(  k_i k_j A^{20}_{ij} - 2 i k_i V^{10}_i  \Big)
+ b_{1} b_{s^2} \Big( 2 i k_i V^{12}_i \Big)
+ b_{2} b_{s^2} \chi^{12} 
+ b_{s^2}^2 \zeta_L \non\\
&~~~
- \frac{1}{2}\alpha_\xi k^2
+ i 2 b_{\nabla^2} \bigg( k_i \frac{\nabla^2}{\Lambda_L^2} U^{(1)}_i \bigg)
+ 2 b_{1} b_{\nabla^2} \Big( \frac{\nabla^2}{\Lambda_L^2} \xi_L \Big)  + \ldots
 \Bigg\}  +  {\rm ``stochastic"},
\end{align}
\vspace{-0.05cm}
where ``stochastic" represents the stochastic contributions to the power spectrum 
which we take to be a scale independent constant up to the order we work at.
The ordering of the terms in this expression follows that for $\xi$ in the main text, so 
that the correspondance is clear. The $d^3q$ integral can be written in spherical polar 
coordinates, and the $\phi_q$ integral trivially gives $2\pi$. 
The $\mathbf{k}\cdot\mathbf{q}$ in the exponent can be written as
$kq\,\mu$, with the standard definitions of those terms. Expanding $k_ik_jA_{ij}$ gives terms 
going as $\mu^0$ and $\mu^2$.  The integral over $\mu$ can then be written as a sum of 
spherical Bessel functions using the identities in Appendix A of Ref.~\cite{VlaWhiAvi15}
resulting finally in sums of 1D integrals:
\begin{align}
  P^{hh} &=(1 - \tfrac{1}{2}\alpha_{\xi} k^2 )P_{\rm zel} + P_{\rm 1-loop}
+ b_{1} P_{b_{1}} 
+ b_{1}^2  P_{b_{1}^2} +b_{2}  P_{b_{2} }
+ b_{1} b_{2} P_{b_{1} b_{2}} 
+ b_{2}^2 P_{b_{2}^2 } 
+ b_{s^2} P_{b_{s^2}} \non\\
&\hspace{-0.4cm} 
+ b_{1} b_{s^2} P_{b_1 b_{s^2}}
+ b_{2} b_{s^2} P_{b_2 b_{s^2}}
+ b_{s^2}^2 P_{(b_{s^2})^2} 
+ 2 b_{\nabla^2} P_{b_{\nabla^2}}
+ 2 b_{1} b_{\nabla^2} P_{b_{1} b_{\nabla^2}}
+  {\rm ``const."}
+\ldots 
\end{align}
\vspace{-0.05cm}
where $P_{\rm zel}$ and $P_{\rm 1-loop}$ are the dark matter contributions, see e.g. Ref.~\cite{VlaWhiAvi15}. 
For additional biasing terms we have terms of the form
\begin{align}
P_{x}= 
4 \pi \int q^2 d q~ e^{ -  \frac{1}{2} k^2 (X_L + Y_L)} \Bigg( f^{(0)}_x(k,q) j_0(qk) + \sum_{n=1}^{\infty} f^{(n)}_x(k,q) \lb \frac{kY_L}{q} \rb^n j_n(qk)
\Bigg), 
\label{eq:PS_real}
\end{align}
where integrands are: 
\begin{center}
\begin{tabular} {  c | c c  }
term & $f^{(0)}_x$  & $f^{(n)}_x$  \\ [0.2em] \hline \hline
$b_1$, &
$- k^2 \big(X^{10}+ Y^{10}\big)$, &
$- k^2 \big(X^{10}+ Y^{10}\big) +2\lb nY^{10} - q U^{10} \rb/Y_L$,  \\[0.5 em]
$b_1^2$, &
$\xi_L - k^2 (U^{10})^2$, &
$\xi_L - k^2 (U^{10})^2 + \lb 2n(U^{10})^2 - q U^{11} \rb /Y_L $,  \\[0.5 em]
$b_2$, &
$- k^2 (U^{10})^2$, &
$- k^2 (U^{10})^2 + \lb 2n(U^{10})^2 - q U^{20} \rb /Y_L $,  \\[0.5 em]
$b_1b_2$, &
$0$, &
$- 2qU^{10}\xi_L/Y_L$,  \\[0.5 em]
$b_2^2$, &
$\frac{1}{2} \xi_L^2$, &
$\frac{1}{2} \xi_L^2$,  \\[0.5 em]
$b_{s^2}$, &
$- k^2 \big(X^{20}+ Y^{20}\big)$, &
$- k^2 \big(X^{20}+ Y^{20}\big) +2 \lb nY^{20} - q V^{10} \rb / Y_L $,  \\[0.5 em]
$b_1b_{s^2}$, &
$0$, &
$- qV^{12}/Y_L$,  \\[0.5 em]
$b_2b_{s^2}$, &
$\chi^{12}$, &
$\chi^{12}$,  \\[0.5 em]
$(b_{s^2})^2$, &
$ \zeta_L$, &
$ \zeta_L$,  \\[0.5 em]
$b_{\nabla^2}$, &
$ 0$, &
$ - q\nabla^2 U^{10}/ (\Lambda_L^2Y_L)  $,  \\[0.5 em]
$b_1 b_{\nabla^2}$, &
$ \nabla^2\xi_L / \Lambda_L^2 $, &
$ \nabla^2\xi_L / \Lambda_L^2 $,
\label{tbl:P_halo_terms}
\end{tabular}
\end{center}
The 1D integrals are Hankel transforms which can be done efficiently using
FFTs \cite{Ham00} as was shown in \cite{VlaSelBal15}.
The translation of the mean pairwise velocity and velocity dispersion terms
is very similar.

\section{Time derivatives}
\label{app:derivs}

We have followed Ref.~\cite{WanReiWhi14} in writing the terms in
Eqs.~(\ref{eqn:v-kernel-m-2nd}, \ref{eqn:sig-kernel-m-2nd}) in terms
of time derivatives.
This is a shorthand, in which e.g.~$\dot{A}_{ij}$ stands for
$\langle\Delta_i\dot{\Delta}_j\rangle$.
If we follow the normal convention and write each of the $A_{ij}$ terms
as $X\delta_{ij}+Y\hat{q}_i\hat{q}_j$, the time derivatives become:
\begin{eqnarray}
  f^{-1} \dot{X}        &=& X^{(11)}+2X^{(22)}+4X^{(13)} \\
  f^{-2}\ddot{X}        &=& X^{(11)}+4X^{(22)}+6X^{(13)} \\
  f^{-1} \dot{X}^{(10)} &=& (3/2)X^{(10)} \\
  f^{-2}\ddot{X}^{(10)} &=& 2X^{(10)}
\end{eqnarray}
and analogously for the $Y$ terms.

The displacement-density correlators behave as
\begin{eqnarray}
  f^{-1}\hat{q}_i\dot{U}_{i}        &=& U^{(1)}+3U^{(3)} \\
  f^{-1}\hat{q}_i\dot{U}_{i}^{(20)} &=& 2U^{(20)} \\
  f^{-1}\hat{q}_i\dot{U}_{i}^{(11)} &=& 2U^{(11)} \\
  f^{-1}\hat{q}_i\dot{V}_{i}^{(10)} &=& 2V_{i}^{(10)} \\
  f^{-1}\hat{q}_i\dot{V}_{i}^{(12)} &=& V_{i}^{(12)} \\
\end{eqnarray}
while $f^{-1}\dot{\Upsilon}_{ij}=\Upsilon_{ij}$ and $f^{-2}\ddot{\Upsilon}_{ij}=\Upsilon_{ij}$.

Finally the $W_{ijn}$ terms go as
\begin{equation}
  f^{-2}\ddot{W}_{ijn} = 2 W^{(112)}_{ijn} +2 W^{(121)}_{inj} + W^{(211)}_{nji}.
\end{equation}
Note that only LPT contributions are considered in this section. 
In addition to the LPT terms we need to consider the counter terms which have 
a time dependent coefficient. This is taken into account in the main text above.

\section{Biasing expansion in the Lagrangian coordinates}
\label{app:bias}

When modeling the clustering of biased tracers to high precision,
the modeling of the biasing can be of equal importance as the higher
order corrections in perturbation theory (and the EFT terms) describing
the nonlinear dynamics.
Phrased in another way, the ``cut-off'' scale associated with the biasing
can be larger than for dynamics, so we might need to keep higher order terms \cite{Sen14,Mirb14,
Ang15,Fujita2016}.
As we move into the future the tightest constraints will increasingly
come from galaxies at higher redshifts (where surveys have the most volume)
which will tend to be brighter and hosted by more massive halos for which
we expect this will be even more true.

In our scheme we follow Ref.~\cite{Mat08a} in assuming local Lagrangian bias,
but extend the model to include a dependence on derivative term
$\nabla^2\delta/\Lambda_L^2$ (with associated Lagrangian scale $\Lambda_L$)
and the tidal shear tensor
(see also \cite{Cat98,Cat00,McDRoy09,Bal12,Cha12,Whi14})
\begin{equation}
  s_{ij}(\vk) = \left( \frac{k_ik_j}{k^2}-\frac{1}{3}\delta_{ij}\right)
  \,\delta(\vk)
\end{equation}
as in Eq.~(\ref{eqn:deltaX}).
We assume that the biasing function is smooth, and can be Taylor expanded.
We keep terms through second order in the field, neglecting the third order
terms shear and local terms\footnote{We note that, by construction, the
field $\psi$ introduced in Ref.~\cite{McDRoy09} vanishes in the initial
conditions.  This term arises (deterministically) due to evolution, so it
does not have a biasing coefficient.}.
Explicitly, we can write for the overdensity of bias tracers in the Lagrangian
coordinates
\begin{align}
\df_X(\vec q) &= c_\df ~ \df_L (\vec q)  + c_{\df^2} \lb \df^2_L (\vec q)  - \la \df^2 \ra \rb + c_{s^2} \lb s^2 (\vec q) - \la s^2 \ra \rb  \nonumber \\ 
& ~~ + c_{\nabla^2} \frac{\nabla_q^2}{\Lambda_L^2} \df_L (\vec q) + \ldots + {\rm ``stochastic"},
\end{align}
where ``stochastic" stands for the stochastic contributions to the overdensity
field of the biased tracers, and we have neglected the third biasing order
terms like $s^3$, $\df s^3$ etc.
Fourier transforming $F$ on $\nabla^2\delta$, $s^2$, ... as well as $\delta$
we need to find the expectation value of exponentials such as
\cite{Mat08a,Mat08b,CLPT}
\begin{equation}
  \left\langle \exp\left[i\left(\vk\cdot\mathbf{\Delta}
  +\lambda_1\delta_1+\lambda_2\delta_2
  +\eta_1\nabla^2\delta_1 + \eta_2\nabla^2\delta_2
  +\zeta_1s_1^2+\zeta_2s_2^2+\cdots\right)\right]\right\rangle .
\end{equation}
This can be evaluated using the cumulant expansion, and the terms depending
on $\eta_i$, $\zeta_i$, etc.~can be Taylor expanded out of the exponential
as is usually done for $\lambda_i$.
As for the EFT counter terms, we treat the terms depending on
$(\nabla^2/\Lambda_L^2)\delta$ and $s^2$ as ``higher order'',
in this case in an expansion of $k$ times the Lagrangian radius
($\sim 1/\Lambda_L$) of the halo.  
When expanding these terms down from the exponential, 
keeping only linear contributions in the exponent,
 we truncate at $\mathcal{O}(P_L^2)$.
We obtain the result given in the Eq.~(\ref{eqn:xi-kernel-m}).
The usual bias terms (due to the local Lagrangian bias) can be expressed in
terms of linear correlation function
$\xi_L=\langle \delta_{1}\delta_{2}\rangle_c$ 
and correlatiors 
\begin{equation}
  U_i^{10}   = \langle \delta_1\Delta_i\rangle_c , \quad
  U_i^{20}   = \langle \delta^2\Delta_i\rangle_c , \quad
  U_i^{11}   = \langle \delta_1\delta_2\Delta_i\rangle_c , \quad
  A_{ij}^{10}= \langle \delta\Delta_i\Delta_j\rangle_c \quad .
\end{equation}
for which expressions are given in Refs.~\cite{Mat08a,Mat08b,CLPT}.
To lowest order the $\nabla^2\delta$ terms introduce only two new
correlators:
\begin{equation}
  \left\langle \nabla^2\delta(\vq_1)\delta(\vq_2)\right\rangle
  \qquad {\rm and}\qquad
  \left\langle \nabla^2\delta(\vq_1)\Psi_i^{(1)}(\vq_2)\right\rangle .
\end{equation}
Calling these $\mathcal{B}_i$ we have
\begin{eqnarray}
  \mathcal{B}_1(q) &\equiv& -\nabla^2\xi_L(q) =
  \int\frac{k^2\,dk}{2\pi^2}\ k^2 P_L(k) W(k;R)\ j_0(kq),  \\
  \mathcal{B}_{2,i}(q) &\equiv&  -\nabla_i\xi_L(q) =
  \hat{q}_i\int\frac{k^3\,dk}{2\pi^2}\ P_L(k)\ j_1(kq)
\end{eqnarray}
where $W$ is a smoothing window which we shall not need explicitly.
Note the similarity of $\mathcal{B}_1$ to the EFT counter term,
which is $k^2P_L(k)$ in Fourier space \cite{VlaWhiAvi15}.
For this reason we shall not include this term in our model, as discussed
in more detail in \S\ref{sec:xir}.

We note that including third order shear terms $s^3$ and $\df s^3$,
interestingly, leads to the trivial contributions in the sense that no
new biasing parameters are needed.
Specifically, we find that terms like
$V_i^{11}=\langle s^2_1\delta_{L1}\Delta_i\rangle_c$,
which involve $s^2$ and $\delta$ at the same point and which would have
given rise to new bias coefficients, vanish.
Thus up to $\mathcal{O}(\delta^4)$ we have
\begin{equation}
  V_i^{10} = \langle s^2\Delta_i\rangle_c , \quad
  V_i^{12} = \langle s^2_1\delta_2\Delta_i\rangle_c , \quad
  \Upsilon_{ij} = \langle s^2\Delta_i\Delta_j\rangle_c
\end{equation}
and
\begin{equation}
  \zeta=\langle s^2_1 s^2_2\rangle_c , \quad
  \chi^{12}= \langle s^2_1\delta_2^2\rangle_c,
\end{equation}
where we have
\begin{align}
\chi^{12} &= \frac{4}{3}\left[ \int\frac{k^2\,dk}{2\pi^2}P(k)\,j_2(kq)\right]^2,
\end{align}
and similarly the expression\footnote{There is a typo in Eq.~(A6) of
Ref.~\cite{Whi14}: the coefficient of the $\mathcal{J}_7\mathcal{J}_9$
should be 4, not 2.} for $\zeta$ can be found in \cite{Whi14}.
The shear correlators introduce new combinations.
Some of these terms can be written as products of integrals of $P_L$, viz
\begin{equation}
  V_i^{12} = 2\hat{q}_i \int\frac{k\,dk}{2\pi^2}\ P_L(k)
  \left[\frac{4}{15}j_1(kq)-\frac{2}{5}j_3(kq)\right]
  \int\frac{k^2\,dk}{2\pi^2}P_L(k)\,j_2(kq), 
\end{equation}
and similarly
\begin{eqnarray}
  \Upsilon_{mn} = 2
  \left\langle s^2_{1,ij}\Psi_{2,m}\right\rangle
  \left\langle s^2_{1,ij}\Psi_{2,n}\right\rangle &=&
  2\left\{
  \delta_{mn}\left(2\mathcal{J}_3^2\right) + \hat{q}_m\hat{q}_n
  \left(3\mathcal{J}_2^2+4\mathcal{J}_2\mathcal{J}_3+\right.\right.
  \nonumber \\
  && \left.\left. 2\mathcal{J}_2\mathcal{J}_4+2\mathcal{J}_3^2+
  4\mathcal{J}_3\mathcal{J}_4+\mathcal{J}_4^2\right)\right\}
\end{eqnarray}
where we have defined \cite{Whi14}
\begin{eqnarray}
  \mathcal{J}_2 &=& \int\frac{k\,dk}{2\pi^2}P_L(k)\left[
  \frac{2}{15}j_1(kq)-\frac{1}{5}j_3(kq)\right] \\
  \mathcal{J}_3 &=& \int\frac{k\,dk}{2\pi^2}P_L(k)\left[
  -\frac{1}{5}j_1(kq)-\frac{1}{5}j_3(kq)\right] \\
  \mathcal{J}_4 &=& \int\frac{k\,dk}{2\pi^2}P_L(k)\,j_3(kq)
\end{eqnarray}
Finally, $V_i^{10}=\langle s^2\Delta_i\rangle_c$ which we can write
\begin{equation}
  V_i^{10} =
  \left\langle s^2(\vq_1)\Psi_i^{(2)}(\vq_2)\right\rangle_c =
  -\frac{2\hat{q}_i}{7}\int\frac{k\,dk}{2\pi^2}Q_{s^2}(k)\,j_1(kq)
\end{equation}
with
\begin{equation}
  Q_{s^2}(k) = \frac{k^3}{4\pi^2}\int dr\ P_L(kr)
  \int dx\ P_L(k\sqrt{y}) Q_{s^2}(r,x),
\end{equation}
where we have followed Ref.~\cite{Mat08a} and written $y=1+r^2-2rx$ and
have defined
\begin{equation}
  Q_{s^2}(r,x) = \frac{r^2(x^2-1)(1-2r^2+4rx-3x^2)}{y^2}.
\end{equation}

\section{Connection to the distribution function formalism}
\label{app:DF}

An alternative formalism for describing redshift-space distortions has been
developed in Refs.~\cite{SelMcD11,Oku12,Oku2012v2, Vla12, Vla2013,Oku2014},
and is known as the distribution function (DF) formalism.
In this formalism the redshift space power spectrum is expanded in terms of
the velocity moments defined as
\begin{equation}
  T^L_\parallel(\vec{x}) = (1+\df(\vec{x})) v^L_\parallel(\vec{x}) ~~ 
  {\rm for}~~L\geq1,~~{\rm and}~~T^0_\parallel (\vec x)=\df(\vec{x}).
\end{equation}
The redshift space power spectrum is then given as 
\begin{equation}
  P^{s}(\vec{k})=\sum_{L,L'}\frac{(-1)^{L'}}{L!L'!}
  \lb \frac{ik_\parallel}{\mathcal{H}}\rb^{L+L'}P_{LL'}(\vec{k}),
\label{eq:ps_rsd_v1}
\end{equation}
where $P_{LL'} = \la T^L_\parallel(\vec{k})\right.\left|T^{*L'}_\parallel \ra'$
are the velocity moment correlators\footnote{For $\la \ldots \ra'$ correlators
we drop the delta function, and it is to be understood that only the `on shell'
momentum is considered}. 

There is a tight connection between the DF formalism and the GSM.
The mean pairwise velocity, $v_{12}$, corresponds to the density-momentum
contribution, $P_{01}$, to $P(k)$ while the pairwise velocity dispersion,
$\sigma_{ij}^2$, corresponds to the $P_{11}$ and $P_{02}$ contributions in
the distribution function approach.
Moreover, expanding the exponent in the GSM and keeping only the terms up
to a given PT order one generates all the disconnected contributions to the
$P_{LL'}$ correlators (up to the same PT order).
In this way, on a basic level, two approaches are equivalent and the
difference comes in through the resummation of the connected contributions.

We can show this connection more explicitly.
The density momentum correlator, $P_{01}$, in the DF formalism can be
directly related to the pairwise velocity (see also \cite{Oku2014}).
We have
\begin{align}
2P_{01,i}(k) = 2\la \df(k)|\big[1+\df(k') \big] \circ u_i(k') \ra'
&=  \int d^3 r\ e^{i \vec k \cdot \vec r} \la \big[1+\df(r) \big] \big[1+\df(r') \big] \Delta u_i(r) \ra' \non\\
& =  \int d^3 r\ e^{i \vec k \cdot \vec r} \left[1+\xi(r)\right] v_{12,i}(r).
\end{align}
or inversely
\begin{equation}
  \left[1+\xi(r)\right] v_{12,i}(r) = 2\int \frac{d^3 k}{(2\pi)^3}
  e^{ - i \vec k \cdot \vec r}\ P_{01,i}(k).
\end{equation}
A similar expression holds for the pairwise velocity dispersion, $\sigma_{12}$,
which we can decompose as
\begin{align}
  \left[1+\xi(r)\right] \sigma_{12,ij}(r) &=
\la \big[1+\df \big] \big[1+\df' \big] \Delta u_i\Delta u_j'  \ra \non\\
&= 2\la \big[1+\df \big] \big[1+\df' \big] u_i' u_j' \ra - 2\la \big[1+\df \big]  u_i \big[1+\df' \big]  u_j' \ra \non\\
&= 2 \lb \xi^{ij}_{02}(0) + \xi^{ij}_{02}(r) - \xi^{ij}_{11}(r) \rb,
\end{align}
where we have introduced correlators
\begin{align}
\xi^{ij}_{02}(0) &= \la \big[1+\df(x) \big] u_i(x) u_j(x) \ra, \non\\
\xi^{ij}_{02}(r) &= \la \df(x) \big[1+\df(x') \big]  u_i(x') u_j(x') \ra, \non\\
\xi^{ij}_{11}(r) &= \la \big[1+\df(x) \big] u_i(x) \big[1+\df(x') \big] u_j(x') \ra.
\end{align}
When Fourier transformed and projected to the line of sight direction $\hat z$,
these give the momentum-momentum $P_{11}$ and density-energy density $P_{02}$
correlators used  in the DF formalism. 

If we restrict our comparison to one-loop in PT, all the higher momentum
correlators $P_{LL'}$
($P_{12},~P_{03},~P_{13},~P_{04}$ and $P_{22}$ at one loop)
can be reduced to the contributions of irreducible components (cumulants)
of terms $P_{00}$, $P_{01}$, $P_{11}$ and $P_{02}$
(or equivalently $\xi$, $v_{12}$ and $\sigma_{12}$) 
\cite{Vla12, Vla2013}.
In streaming models these higher momentum correlators (at one loop)
correspond to products of cumulants obtained from expansion of the
exponent in Eq.~\eqref{eq:gsm}.
At higher PT orders these momentum correlators start to collect the
nontrivial (irreducible) contributions from the cumulants $C$, $C^i$
and $C^{ij}$ of Appendix \ref{app:GSM}.

\bibliographystyle{JHEP}
\bibliography{ms}
\end{document}